\documentclass[11pt]{article}
\usepackage{jheppub}
\pdfoutput=1
\usepackage{epstopdf}

\usepackage{jheppub}
\usepackage{mathrsfs}
\usepackage{psfrag}
\usepackage{color}
\usepackage{hyperref}

\usepackage{slashed}
\usepackage{feynmp-auto}
\usepackage{simplewick}
\usepackage{cancel}

\usepackage{todonotes}

\usepackage{amsmath,bbm,array,amsfonts,graphicx,wrapfig,arydshln,lscape,float,multirow,longtable,rotating,makecell}
\usepackage{url}

\DeclareMathAlphabet\mathbfcal{OMS}{cmsy}{b}{n}

\newcommand{\la}{\langle}
\newcommand{\ra}{\rangle}

\title{Unwinding the Amplituhedron in Binary}

\author{Nima Arkani-Hamed,$^1$}
\author{Hugh Thomas,$^{2}$}
\author{Jaroslav Trnka$^{3}$}

\affiliation{$^1$ School of Natural Sciences, Institute for Advanced Study, Princeton, NJ 08540, USA}

\affiliation{$^2$ LaCIM, D\'epartement de Math\'ematiques, Universit\'e du Qu\'ebec \`a Montr\'eal, 
Montr\'eal,  QC, Canada}

\affiliation{$^3$ Center for Quantum Mathematics and Physics (QMAP), \\ Department of Physics, University of California, Davis, CA 95616, USA}

\emailAdd{arkani@ias.edu, hugh.ross.thomas@gmail.com, trnka@ucdavis.edu}

\abstract{
We present new, fundamentally combinatorial and topological characterizations of the amplituhedron. Upon projecting external data through the amplituhedron,  the resulting configuration of points has a specified (and maximal)  generalized ``winding number". Equivalently, the amplituhedron can be fully described in binary: canonical projections of the geometry down to one dimension have a specified (and maximal) number of ``sign flips" of the projected data. The locality and unitarity of scattering amplitudes are easily derived as elementary consequences of this binary code. Minimal winding  defines a natural ``dual" of the amplituhedron. This picture gives us an avatar of the amplituhedron purely in the configuration space of points in vector space (momentum-twistor space in the physics),  a new interpretation of the canonical amplituhedron form, and a direct bosonic understanding of the scattering super-amplitude in planar ${\cal N}=4$ SYM as a differential form on the space of physical kinematical data.}

\preprint{
\begin{flushright}\end{flushright}
}

\begin{document}

\maketitle

\newpage

\section{The Amplituhedron}

Recent years have revealed a fascinating and unexpected connection between the basic physics of particle scattering amplitudes and new mathematical structures in ``positive geometry" \cite{ArkaniHamed:2012nw, L1,L2,P}. In the context of ${\cal N}=4$ super-Yang-Mills theory in the planar limit, the Amplituhedron \cite{Arkani-Hamed:2013jha} provides an autonomous definition of scattering amplitudes in purely geometric terms, with no reference to quantum-mechanical evolution in space-time. The principles of locality and unitarity are moved from their primary position in the usual formulation of quantum field theory, to derivative notions emerging hand-in-hand from the positive geometry. This physics and mathematics has  been explored from a variety of perspectives in the past few years (see e.g. \cite{Arkani-Hamed:2013kca,Franco:2014csa,Bai:2014cna,
Lam:2014jda,Arkani-Hamed:2014dca,Galloni:2016iuj,Ferro:2015grk, Ferro:2016zmx,Ferro:2016ptt,
Karp:2016uax,Bai:2015qoa}), and a systematic mathematical exploration of the notion of ``positive geometries" has recently been initiated in \cite{Arkani-Hamed:2017tmz}.

The amplituhedron is a simple generalization of the notion of plane polygons into the Grassmannian. Thinking projectively, the vertices of a convex $n$-polygon can be represented as 3-vectors ${\cal Z}_a^I$ for $a=1, \cdots, n$ and $I=1, \cdots,3$. The convexity is reflected by positivity of minors $[{\cal Z}_a {\cal Z}_b {\cal Z}_c]>0$ for $a<b<c$. Then the interior of the polygon can be thought of as all the points $Y^I$ which are in the convex hull of the ${\cal Z}_a^I$, i.e. all $Y^I$ of the form $Y^I = c_a {\cal Z}_a^I$ with $c_a>0$. The (tree) amplituhedron ${\cal A}_{m,k,n}$ lives in the space of $k$-planes $Y$ in $(k+m)$ dimensions. We have external data ${\cal Z}_a^I$, for $I=1, \cdots, (k+m)$. We think of $Y$ as being the span of $k$ vectors $Y_\alpha^I$ for $\alpha = 1, \cdots k$. We then consider all the $Y_\alpha^I$ of the form
\begin{equation}
Y_\alpha^I = C_{\alpha a} {\cal Z}_a^I
\end{equation}
where the fixed external data ${\cal Z}_a$
is ``positive" in the sense of the ``positive Grassmannian", and we vary over $C_{\alpha a}$ that are also positive in the same sense: 
\begin{equation}
[{\cal Z}_{a_1} \cdots {\cal Z}_{a_{k+m}}] > 0 \,\,\, {\rm for} \,\,\, a_1< \cdots < a_{k+m}, \,\,\, [C_{a_1} \cdots C_{a_k}]>0 \,\,\, {\rm for} \,\,\, a_1 < \cdots < a_k 
\end{equation}
and the simple idea of ``hiding particles" gives a natural extension of this geometry to the ``all-loop" amplituhedron. 
This definition needs an ordering $(1,2,\cdots,n)$ for the external data, but the notion of positivity allows for a ``twisted" cyclic symmetry. If the minors of $C_{\alpha a}$ are positive, so are the minors of a new matrix where $C_{\alpha 1} \to C_{\alpha 2}$, $C_{\alpha 2} \to C_{\alpha 3}, \cdots, C_{\alpha n} \to (-1)^{k-1} C_{\alpha 1}$. The same is true for the ${\cal Z}_a$. Note that if $m$ is even, $(-1)^{k-1} \times (-1)^{k+m-1} = 1$ and so the amplituhedron itself is invariant under an untwisted cyclic symmetry, while for $m$ odd the ordering is reflected in the amplituhedron geometry as well. 

(We break slightly with earlier notation in the literature where the external data is referred to as non-caligraphic $Z_a^I$ since we are reserving $Z_a^I$ for something else we will introduce shortly, and which will make a more ubiquitous appearance in this paper: the data we get after projecting the ${\cal Z}_a$ through $Y$. Also, strong emphasis on positivity associated with the the positivity of the $C_{\alpha a}$ matrix, which played a starring role in the story of on-shell diagrams, and was already ``demoted" to playing an equal role with the positivity of external ${\cal Z}_a$ data in the first description of the amplituhedron, is essentially entirely absent in our new picture. Therefore, no familiarity with the non-trivial aspects of the positive Grassmannian is assumed in what follows. The few ``positive properties" we will use will be introduced in a self-contained way as needed).  

Note that this description of the amplituhedron is highly redundant. This is clear already for the polygon, since the space of the coefficients $c_a$ is (projectively) $(n-1)$ dimensional, while the space of $Y$'s in the polygon is obviously only 2-dimensional. More generally the space of the $C_{\alpha a}$ is $k (n-k)$ dimensional which (since $n \geq (k+m)$) is always larger than $k \times m$ which is the dimensionality of the tree amplituhedron. Concretely, this means that if we are given some $Y$, we can't easily check whether or not it is in the amplituhedron. 
We would like a different description of the amplituhedron, one which can be used to directly check whether or not a given $Y$ is in the amplituhedron. 

This is what we will do in this paper. We will give a radically different, more invariant and intrinsic definition of the amplituhedron, which is essentially entirely combinatorial/topological in nature.  While we do not yet have a complete proof of the equivalence of this new definition with the usual one, we have checked the equivalence numerically in a large number of examples, and will also provide proofs in a number of special cases.  This new picture opens up new avenues of investigation into the structure of the amplituhedron, and also suggests a striking new picture of scattering super-amplitudes in ${\cal N}=4$ SYM, {\it directly} as certain differential forms on the (momentum-twistor) space of external kinematical data. We will briefly touch on a number of these points, deferring more detailed investigations to future work.

\section{Projecting Through $Y$}

We have posed a concrete question which motivates the search for a new definition of the amplituhedron: given some $Y$, how can we check whether it is inside the amplituhedron? 
Now for general convex polytopes, there is a standard answer to this question. Indeed, polytopes can be defined in two different ways. The first is ``vertex-centered": given a collection of points ${\cal Z}^I_a$, the polytope is defined as the convex hull of these points. This is the $``Y = c_a{\cal Z}_a"$ description, which we directly generalize with the conventional definition of the amplituhedron. But there is also a second, ``face-centered" description of the polytope. Here we cut out the polytope by a collection of inequalities associated with the facets ${\cal W}_{I,i}$ of the polytope, i.e. by imposing the inequalities $[\, Y {\cal W}_i  ]\, \geq 0$. 

Can we extend this simple picture to the amplituhedron? We certainly know all the co-dimension one boundaries of the amplituhedron. For instance for $m=2$, this corresponds to $[\, Y i i+1 ]\, \to 0$; for $m=4$, $[\, Y i i+1 j j+1 ]\, \to 0$ etc. (Note that here, and sometimes in what follows, when it will not cause confusion, we write $i$ for ${\cal Z}_i$.)  So it is natural to ask, for instance for $m=2$: is the amplituhedron characterized by $[\, Y i i + 1 ]\, \geq 0$?

The answer is easily seen to be ``no". The obstruction is a familiar one from the usual story of the positive Grassmannian, and can be seen in the first non-trivial case of $k=2,m=2,n=4$ where the amplituhedron corresponds to the simplest positive Grassmannian $G_+(2,4)$. The inequalities associated with the codimension one boundaries are $[\, Y 1 2 ]\, , [\, Y 2 3]\,  , [\, Y 3 4 ]\, , [\, Y 1 4 ]\,$ all $>0$. But then the Plucker relations tell us that 
\begin{equation}
[\,Y 1 3 ]\, [\, Y 2 4 ]\, = [\, Y 1 2 ]\, [\, Y 3 4 ]\, + [\, Y 2 3 ]\, [\, Y 1 4 ]\,
\end{equation}
The right hand side is positive when the boundary inequalities are satisfied, but this doesn't fix the signs of $[\, Y 1 3 ]\,, [\, Y 24 ]\,$, which can be either both positive or both negative. The amplituhedron demands the choice where $[\,Y 1 3 ]\, , [\, Y 2 4 ]\, < 0$, so we see that, unlike for polygons, the boundary inequalities are insufficient to define the space.

Let us start by defining the elementary notion of ``projection", which we will use repeatedly in the rest of this paper. 
Given an $N$-dimensional vector space $\mathbfcal {V}$, there is an obvious notion of projection through some fixed vector ${\cal V}_*$ to get an $(N-1)$ dimensional vector space. The vectors in the new space are 
just the equivalence classes $[{\cal V}] = \{{\cal V}  + \alpha {\cal V}_*\mid {\cal V} \in \mathbfcal {V}\}$. Algebraically, we can always do a $GL(N)$ transformation to put ${\cal V}_*$ in the form ${\cal V}_* = (0, \cdots, 0, 1)$. A vector ${\cal V}$ is then of the form ${\cal  V}  = (V_1, \cdots V_{N-1}, \xi)$, and we can associate the projected $(N-1)$ dimensional vector $[{\cal V}]$ with $V=(V_1, \cdots, V_{N-1})$. Note that those $GL(N)$ transformations that leave ${\cal V}_*$ invariant simply act as $GL(N-1)$ transformations on the projected vectors $V$. The vector ${\cal V}_*$ itself is projected to the origin in the new space. The projection also has an obvious geometric description. We choose some $(N-1)$ dimensional plane passing through the origin and not containing $\cal V_*$; then given any vector ${\cal V}$, we translate it in the direction parallel to the vector ${\cal V}_*$ till it intersects that plane, giving the point $V$. Different choices of the $(N-1)$ plane act as $GL(N-1)$ transformations on $V$. We can similarly start from an $N$ dimensional space and project through a $K$-plane to get to an $(N-K)$ dimensional space. 

$$
\includegraphics[scale=.45]{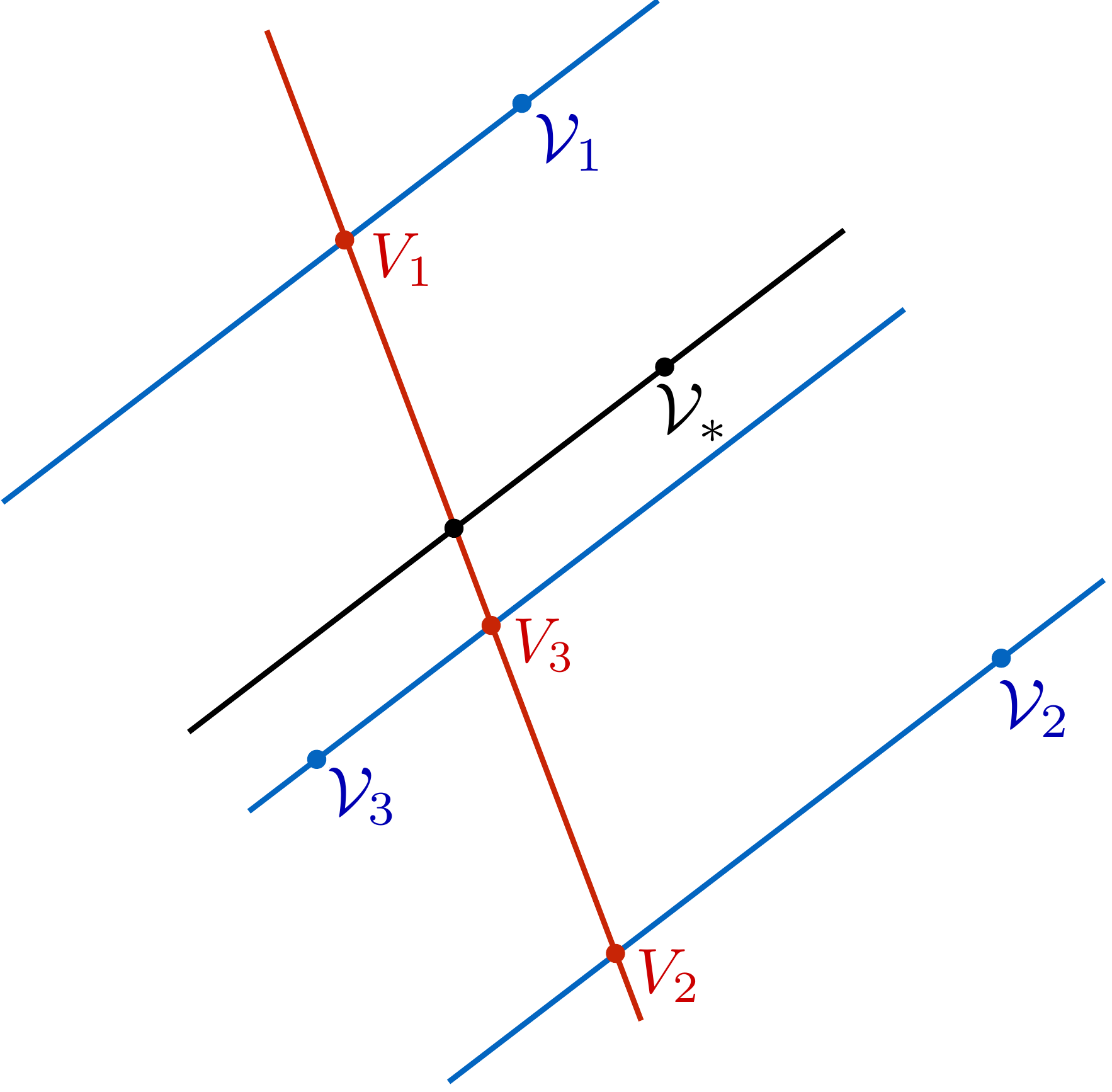}
$$

We will be interested in taking the configuration of $(k+m)$ dimensional vectors ${\cal Z}_a$, and projecting them through the $k$-plane $Y$ to get a configuration of $m$-dimensional vectors $Z_a$. To  repeat the general construction, we can always do a $GL(k+m)$ transformation to put the $k \times (k+m)$ matrix $Y$ in the form $Y = \left({\mathbf 0}_{k \times m} |{\mathbf 1}_{k \times k} \right)$. 
Then the ${\cal Z}_a = (Z_{a} | \xi_{a})$. The $GL(k+m)$ transformations that leave $Y$ invariant act as $GL(m)$ transformation on the $Z_a$. There is also an obvious relationship between the antisymmetric brackets in $(k+m)$ and $m$ dimensions. Representing $Y$ as the span of $k$ vectors $Y_{\alpha = 1, \cdots, k}$, 
\begin{equation}
\langle Z_{a_1} \cdots Z_{a_m} \rangle = [\, (Y_1 \cdots Y_k) {\cal Z}_{a_1} \cdots {\cal Z}_{a_m}]\, \equiv [\, Y {\cal Z}_{a_1} \cdots {\cal Z}_{a_m} ]\,
\end{equation}

We will spend the rest of this section examining what these projections look like for the cases of $m=2$ and $m=1$, and see how the amplituhedron is specified by the elementary notions of ``winding" and ``crossings" in these two cases; this will motivate the analogous definitions for general even and odd $m$ we give in subsequent sections. 

To start with the case $m=2$, we will project the external $(k+2)$-dimensional ${\cal Z}_a$ data through a $k$-plane $Y$ and draw the resulting configuration of $Z_a$ vectors in 2 dimensions. We begin with the case $k=2,n=4$, where the configurations come in two shapes: 

$$
\includegraphics[scale=.5]{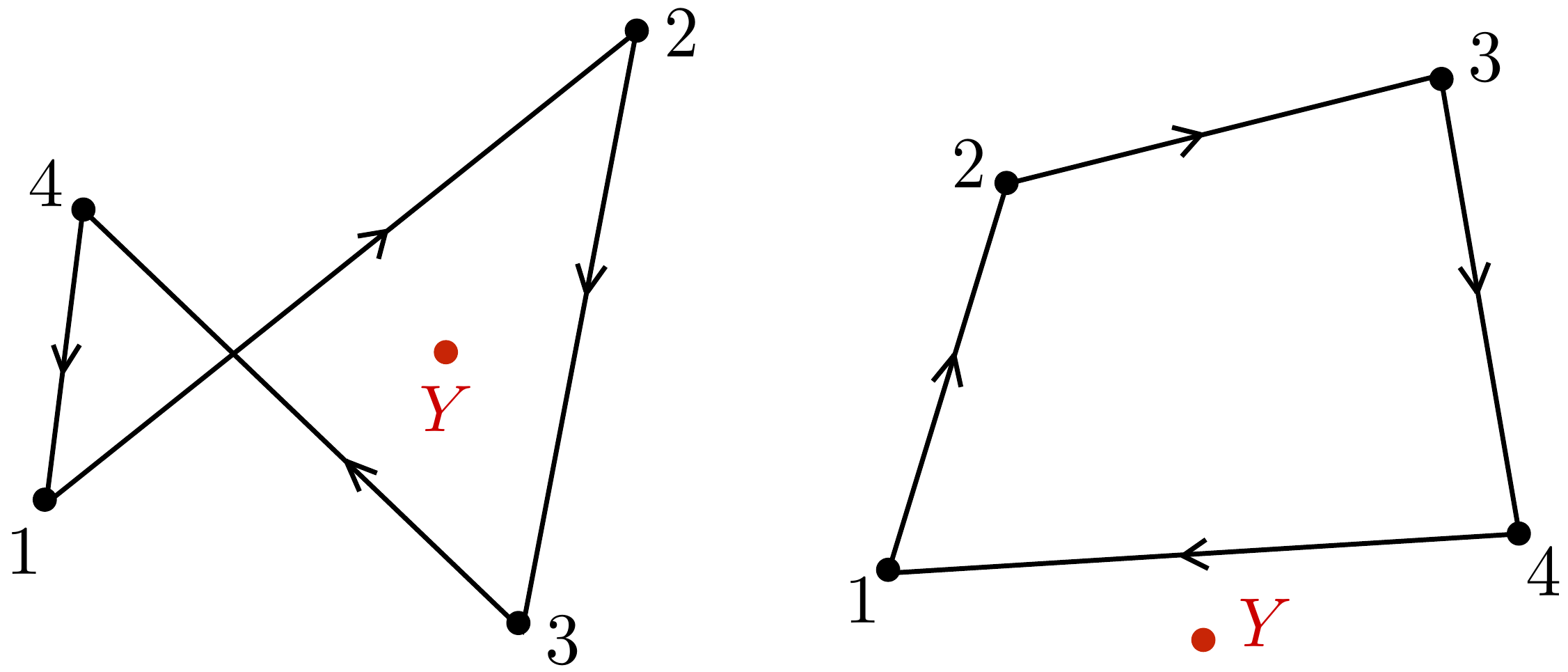}
$$

Since we are projecting through $Y$, all of $Y$ is mapped to the origin in this picture. Note that in both pictures we have $\langle i i+1 \rangle = [\,  Y i i+1 ]\,> 0$, so that all the segments $12,23,34,14$ ``wind around $Y$" with the same orientation.
(Note that we have $[Y 14]>0$, and not $[Y 4 1]>0$. This is a reflection of the twisted cyclic symmetry for even $k$. For odd $k$ we would have $[Y n 1]>0$.) In the first picture, though, the line segments $13,24$ wind oppositely and we have $\langle 1 3 \rangle, \langle 2 4 \rangle < 0$, while in the second configuration they wind in the same direction and $\langle 13 \rangle, \langle 2 4 \rangle > 0$. We can see that to characterize the $Y$'s in the amplituhedron, we must require not only the correct orientation of the segments (i.e. $\langle i i+1 \rangle = [\, Y i i+1 ]\, > 0$), but also that the closed path $(12),(23),(34),(41)$ has a {\it winding number} of 1 around $Y$. 

We can easily repeat this exercise for general $k$.  For the minimal value of $n=k+2$, the signs of all of the $[Y a b ]$ are fixed, and we show the pictures for $k=1, \cdots, 4$ below: 

$$
\includegraphics[scale=.5]{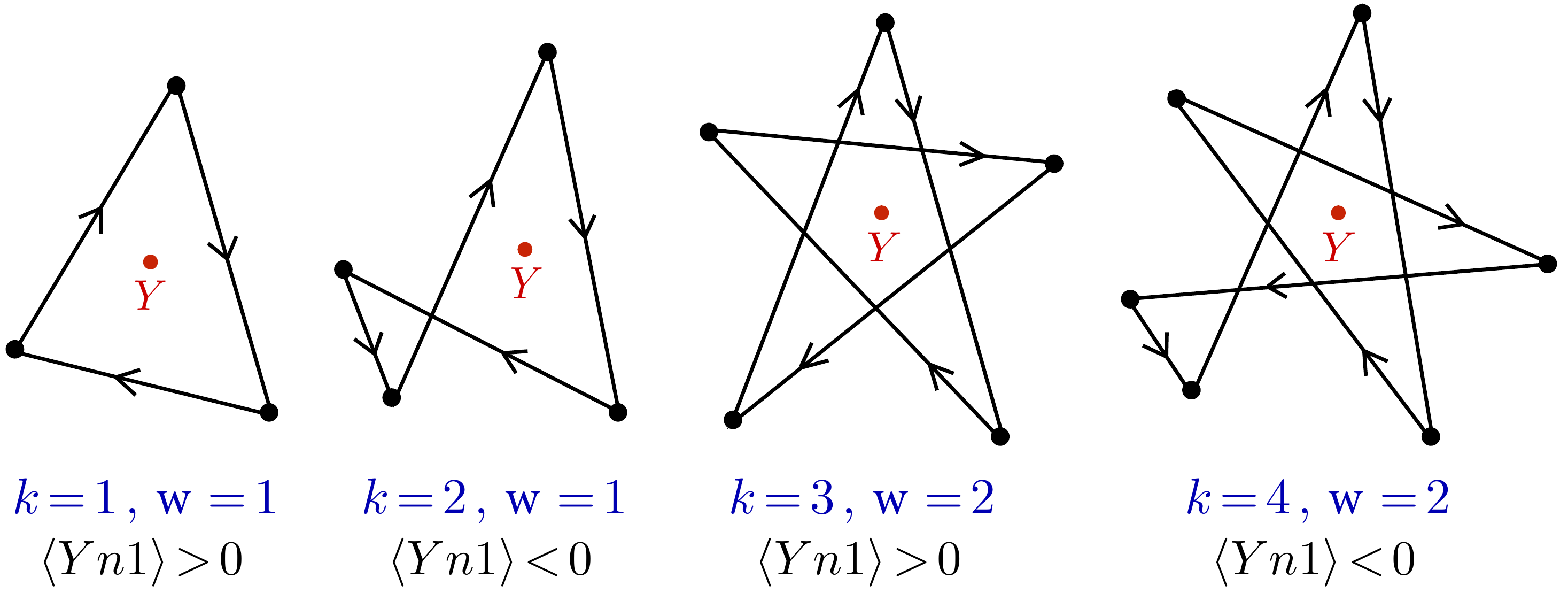}
$$

Using the fact that we are in the $n=k+2$ case, where the signs of all $[Yab]$ are fixed, we see that $Y$ being in the amplituhedron is characterized by the winding number of the path $(12),(23), \cdots, (n1)$. The necessary winding is $w=(k+1)/2$ when $k$ is odd, and $w=(k/2)$ when $k$ is even. Note that these two cases are simply distinguished by $(-1)^{k-1}$ factors associated with the twisted cyclic symmetry, which tells us that $\langle n 1 \rangle = [\, Y n 1 ]\, >0$ for $k$ odd and $\langle  n 1 \rangle = [\, Y n 1 ]\,< 0$ for $k$ even. As we will argue, this picture works for all $n$: $Y$ is in the amplituhedron if and only if $\langle i i+1 \rangle > 0$, and the path $(12),(23), \cdots, (n1)$ has winding number $w=\lfloor \frac{k+1}{2} \rfloor$. 

What happens for $m=1$? Here the only obvious co-dimension one boundary  inequalities correspond to $(-1)^{k} [\, Y 1 ]\, >0, [\, Y n ]\, > 0$, which can't cut out the amplituhedron (for one thing they can't even distinguish between different $k$'s!). But let us follow the same logic as for $m=2$, and ask what the picture looks like after we project through $Y$. Here the final space is even simpler---it is only $1$-dimensional! Clearly we can't be talking about the notion of ``winding number" as we did for $m=2$, but we can do something even more primitive: we can look at the number of times the path $(12),(23), \cdots, (n-1 n)$ jumps over $Y$ (again mapped to the origin), or, equivalently, we can count the number of sign flips in the sequence $\{\langle 1 \rangle, \cdots, \langle n \rangle\}$. Looking again at the case of minimal $n=(k+1)$ reveals the pattern we are looking for: 

$$
\includegraphics[scale=.57]{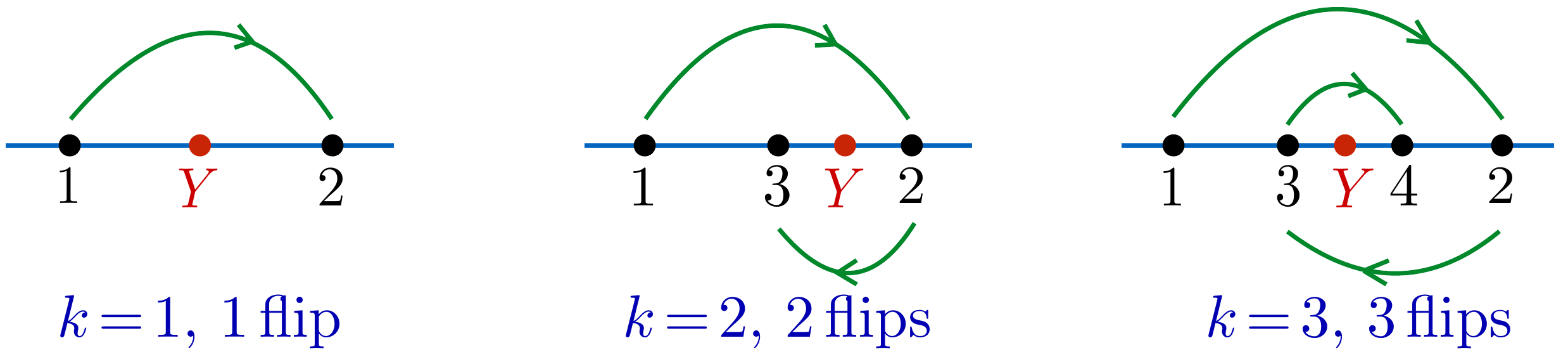}
$$

Again this extends for general $n$: $Y$ is in the $m=1$ amplituhedron if the sequence $\{\langle 1 \rangle,  \cdots, \langle  n \rangle\}=\{[\, Y 1 ]\,, \cdots, [\, Y n ]\,\}$ has exactly $k$ sign flips. 

It is interesting to note that a natural relationship between the ``winding" and ``flip" pictures. Consider an $m=2$ configuration. Then, if we project through e.g.   the point $Z_1$, to go down to one dimension, the resulting configuration of the projected $Z_2,\cdots,Z_n$  has the sign flip pattern compatible with the $m=1$ amplituhedron, i.e. it has precisely $k$ sign flips.

$$
\includegraphics[scale=.52]{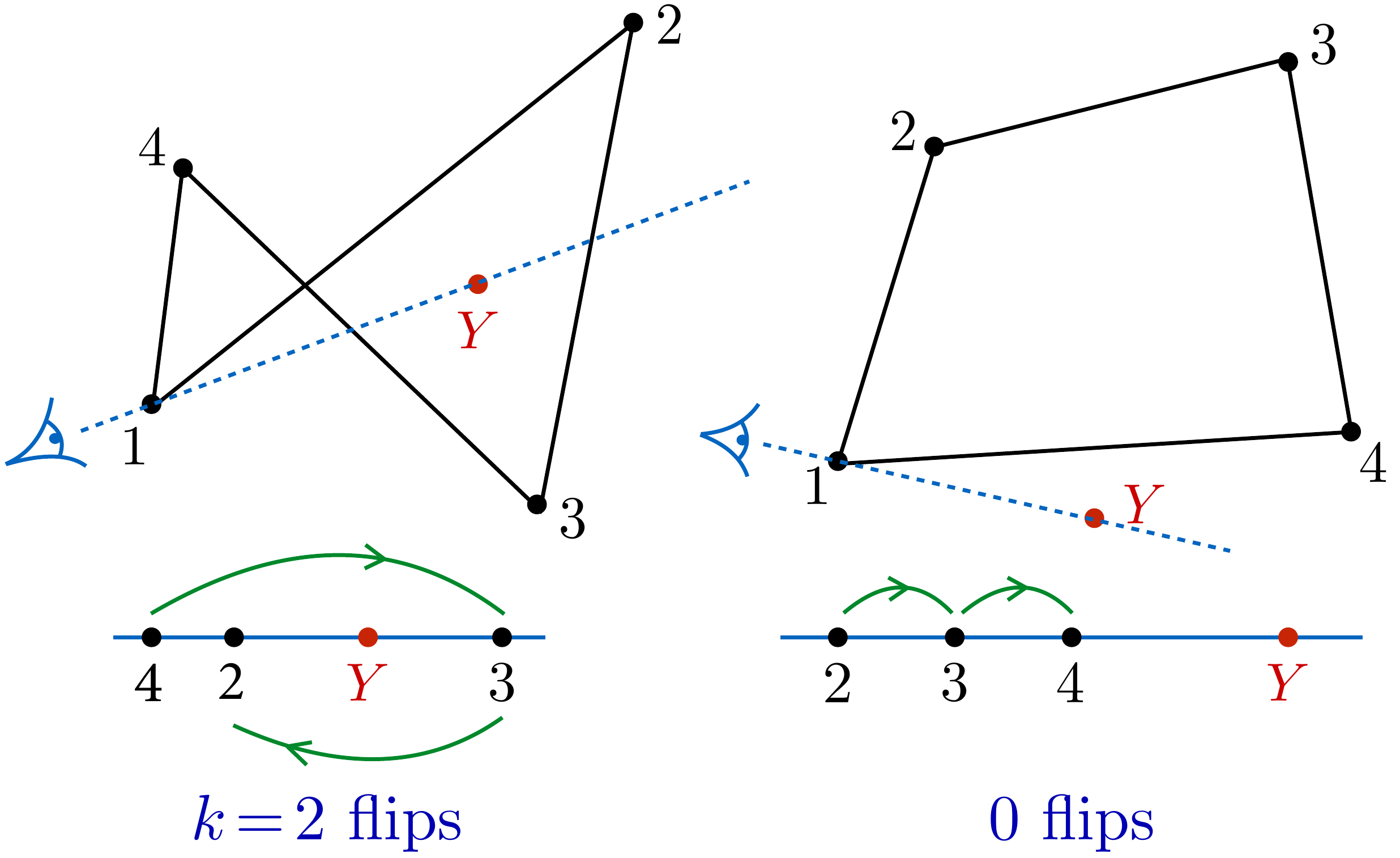}
$$

Indeed, projecting down to one dimension from two dimension gives us a way to characterize the points in the amplituhedron without explicitly assuming that we are on the right side of the boundaries!  We can instead simply demand that we get the correct sign flip pattern upon projecting through {\it each} of the vertices $Z_a$, (and as always appropriately including the factors of $(-1)^{k-1}$ for the twisted cyclic symmetry).  In the righthand figure below, $Y$ is in the amplituhedron, which can be verified either because it is on the right side of the boundaries and has the correct winding, or because in each of the projected one-dimensional pictures, the number of flips equals $k$.  In the lefthand figure, $Y$ is not in the amplituhedron, which can be verified either by observing that it is on the wrong side of the $(34)$ boundary, or that the number of sign-flips in the projected one-dimensional pictures is not always equal to $k$.   

$$
\hspace{-0.5cm}\includegraphics[scale=.45]{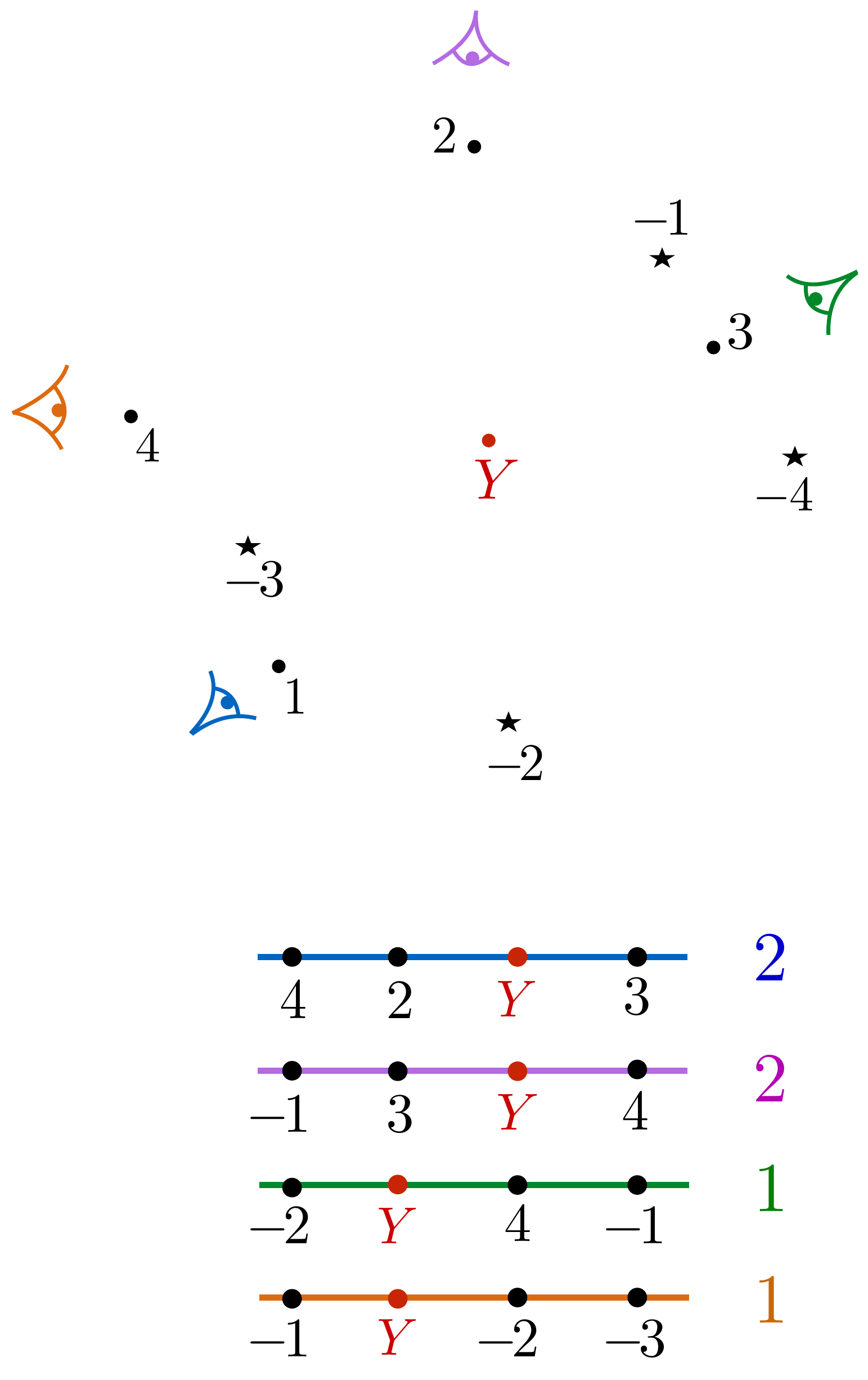}\hspace{2cm}
\includegraphics[scale=.45]{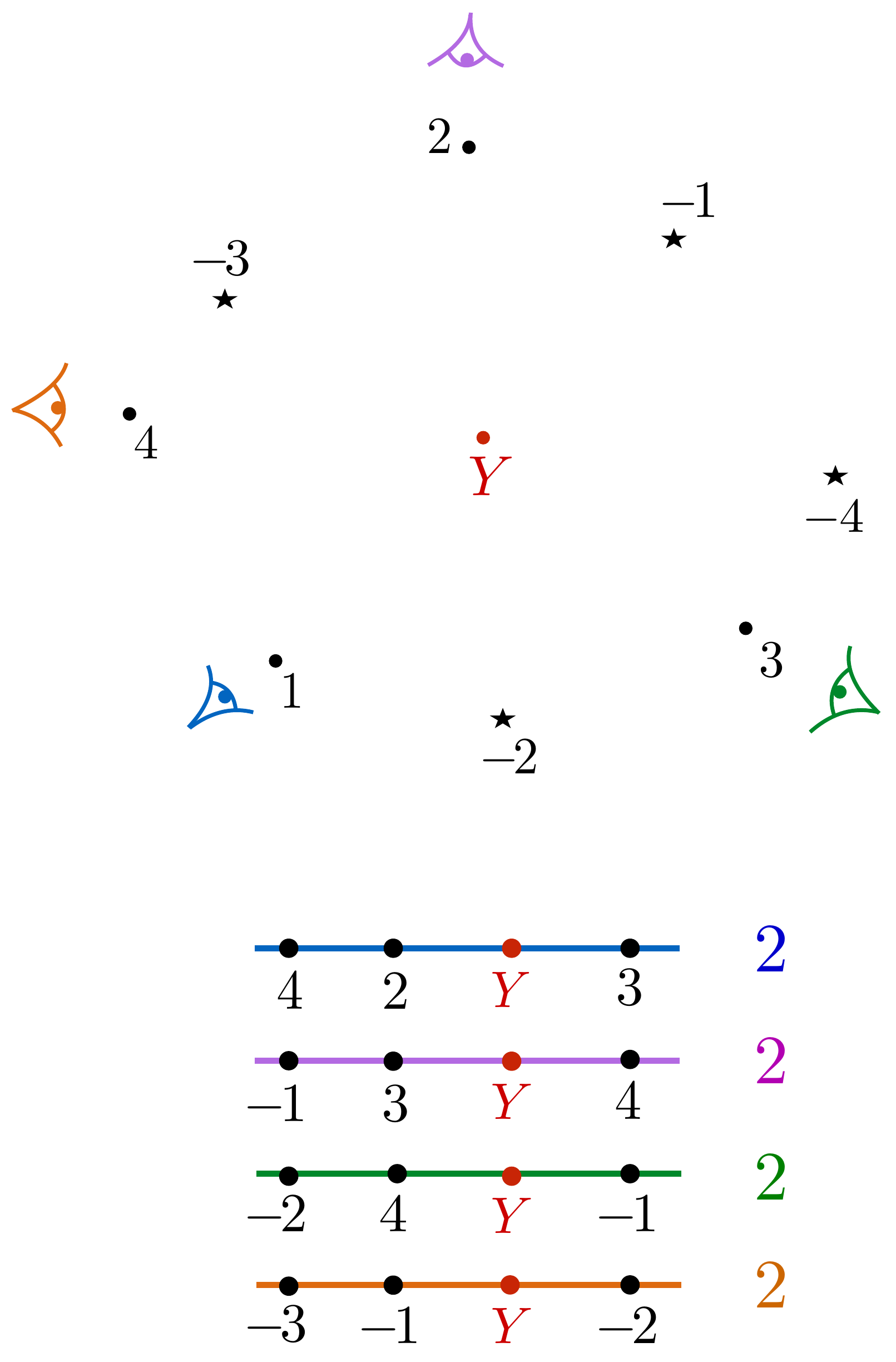}
$$

\section{Winding}

Having motivated our approach to characterizing the amplituhedron with simple examples, we now give a more systematic account starting with the case of even $m$, where we will use a generalized notion of ``winding number". Let us first precisely define what we mean by winding number (again for even $m$); since this is a general topological notion we  will do this for a completely generic configuration of $Z_a$. 

Start with $m=2$. We can count the winding number by asking whether or not a vector pointed in some direction $Z_*$ will intersect the interior of a given boundary $(i i+1)$. 
This means that some positive multiple of $Z_*$ should be expressible as a {\it positive} linear combination of $Z_i$ and $Z_{i+1}$, i.e. that we should be able to express
\begin{equation}
x_* Z_* = x_i Z_i + x_{i+1} Z_{i+1}\,\,\,{\rm with} \,\,\, x_*,x_i,x_{i+1} > 0
\end{equation}

This tells us that a vector in the direction $Z_*$ intersects the boundary $(i i+1)$ if and only if $\frac{\langle  Z_* Z_{i} \rangle}{\langle  Z_i Z_{i+1} \rangle} < 0, \frac{\langle  Z_* Z_{i+1}\rangle}{\langle  Z_i Z_{i+1} \rangle} > 0$. This leads us to define
\begin{equation}
w_i(Z_*) = \left\{\begin{array}{c c} + 1 & \textrm{ if sgn} \{\langle Z_i Z_{i+1} \rangle,
\langle  Z_* Z_{i} \rangle, \langle Z_* Z_{i+1} \rangle \}= \{+,-,+\} \, {\rm or} \{-,+,-\} \\ 0 &  {\rm otherwise} \end{array} \right.
\end{equation}

$$
\includegraphics[scale=.6]{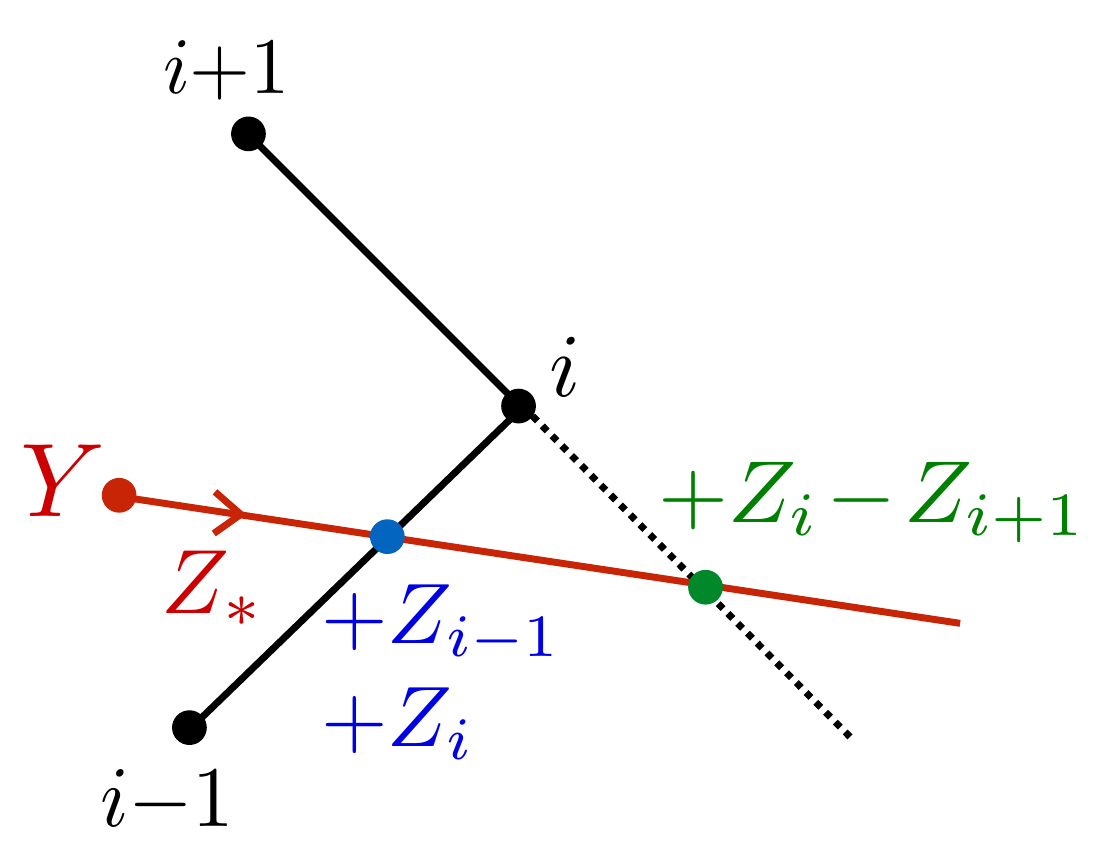}
$$

Then we define the total winding number to sum all the boundaries that are hit in this way, with a factor of $+1$ when they are oriented as $\langle  i i+1 \rangle>0$, and $(-1)$ when $\langle  i i+1 \rangle <0$:
\begin{equation}
w_{m=2} = \sum_i {\rm sgn}(\langle Z_i Z_{i+1} \rangle) \times w_i(Z_*)
\end{equation}
Note that in our applications, where we demand that $\langle i i+1 \rangle > 0$ with the twisted cyclic symmetry, we only pick up a minus sign for the boundaries $(n1)$, and only when $k$ is even. 

The total winding number does not depend on $Z_*$. This is both intuitively obvious and easy to prove. As we change $Z_*$ smoothly, the $w_i$ will not change till the line pointing in the direction of $Z_*$ is hitting the boundary of some interval $(ii+1)$. Let's follow what happens as we start with some boundary $(i i+1)$ that is hit---where we can expand $x_* Z_* = x_i Z_i + x_{i+1} Z_{i+1}$ with $x_*,x_i,x_{i+1} > 0$, and move  $x_{i+1}$ to be very slightly positive, then zero, then slightly negative. Right on the boundary where $x_{i+1} \to 0$, $Z_*$ is obviously also on the boundary of the different interval $(i-1 i)$, so it is natural to ask about whether or this interval is also hit. For small $x_{i+1}$, 
\begin{equation}
  \frac{\langle * i-1 \rangle}{\langle i-1 i \rangle} = \frac{-x_i}{x_*},\qquad \frac{\langle * i \rangle}{\langle i-1 i \rangle} = \frac{-x_{i+1} \langle i i+1 \rangle}{x_*\langle i-1 i \rangle}.
\end{equation} 

Thus if the signs of $\langle  i i+1 \rangle$ and $\langle  i-1 i \rangle$ are the same, then when $x_{i+1}$ is slightly positive we intersect $(ii+1)$ but not $(i-1 i)$, and when we pass through to $x_{i+1}$ slightly negative we no longer intersect $(i i+1)$ but {\it do} intersect $(i-1 i)$. Thus $w_{i} + w_{i-1} = 1$ for both signs of $x_{i+1}$ and the total winding number doesn't change. 
On the other hand when the signs of $\langle  i-1 i \rangle$ and $\langle  i i+1 \rangle$ are opposite, then when $x_{i+1}$ is slightly positive {\it both} intervals are hit, while when $x_{i+1}$ crosses to be slightly negative {\it neither} of the intervals is hit. Thus the sum of the contributions to the winding from $(i-1 i)$ and $(i i+1)$ are zero for both signs of $x_{i+1}$ and again the total winding number doesn't change. 

$$
\includegraphics[scale=.55]{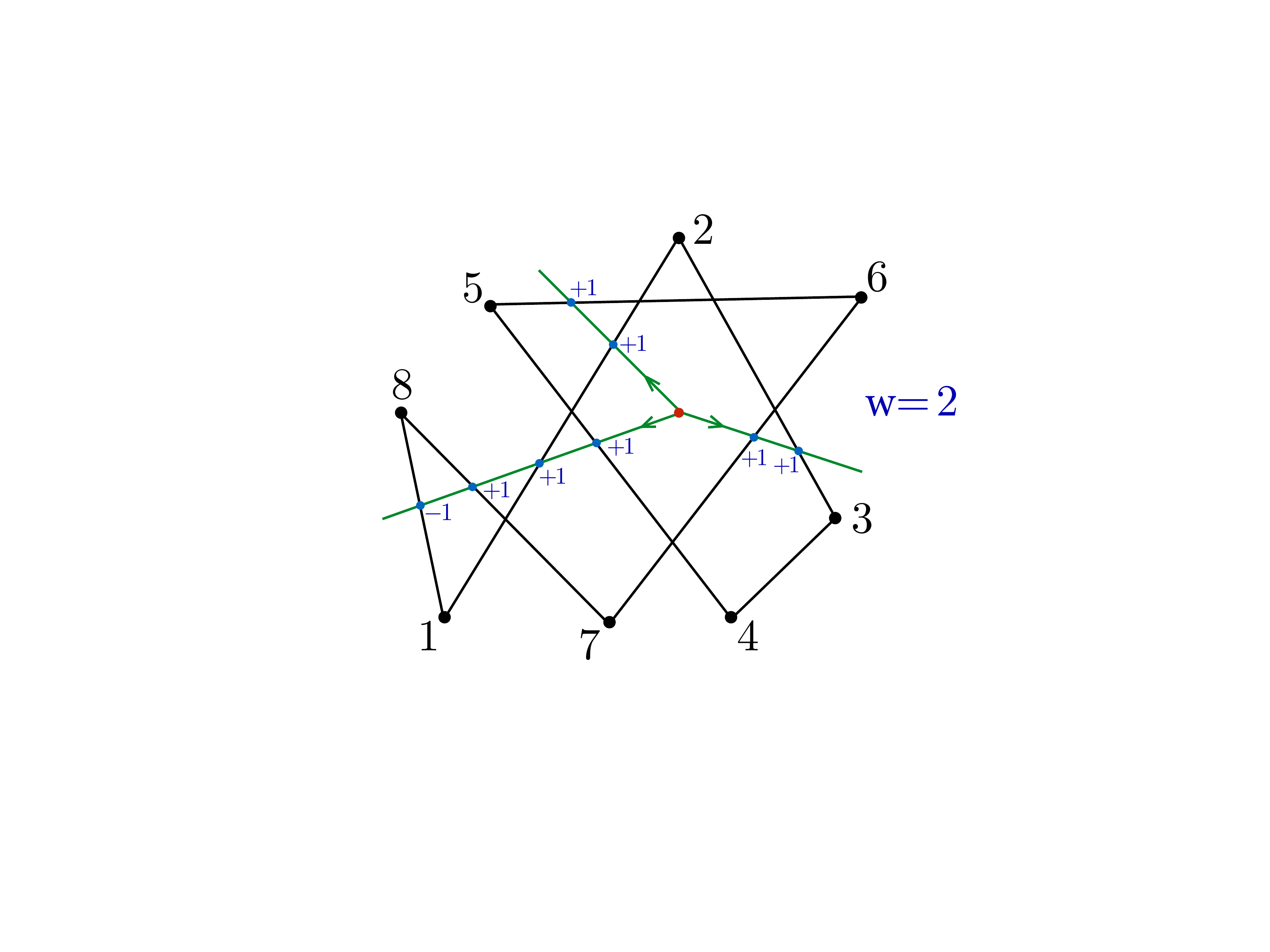}
$$

We can immediately extend to $m=4$. Now, projecting through $Y$ produces points $Z_a$ in a four-dimensional vector space.  In four dimensions, it is not meaningful to talk about the winding of a curve around the origin.  The obvious generalization is to ask about the winding of some topological 3-sphere around the origin instead.  There is a 3-sphere naturally present in the story: the piecewise linear sphere formed from the simplices $(i i+1 j j+1)$.  To understand this winding very concretely, we ask whether a vector in the direction $Z_*$ intersects a given boundary $(i i+1 j j+1)$, which demands that 
\begin{equation}
x_* Z_* = x_i Z_i + x_{i+1} Z_{i+1} + x_j Z_j + x_{j+1} Z_{j+1} \, {\rm with} \, x_*,x_i,x_{i+1},x_j,x_{j+1} > 0
\end{equation}
This tells us that a vector in the direction $Z_*$ intersects the boundary $(i i+1jj+1)$ if and only if $\frac{\langle  Z_* Z_iZ_{i+1} Z_j\rangle}{\langle  Z_i Z_{i+1} Z_j Z_{j+1} \rangle} < 0, \frac{\langle  Z_* Z_{i}Z_{i+1}  Z_{j+1} \rangle}
{\langle  Z_i Z_{i+1} Z_j Z_{j+1} \rangle} > 0, \frac{\langle  Z_* Z_i Z_j Z_{j+1} \rangle}{\langle  Z_i Z_{i+1} Z_j Z_{j+1} \rangle} < 0, \frac{\langle  Z_*  Z_{i+1} Z_{j}Z_{j+1}\rangle}{\langle  Z_i Z_{i+1} Z_j Z_{j+1} \rangle} > 0 $. Again this leads us to define 
\begin{equation}
w_{i,j}(Z_*) =   \left\{\begin{array}{c c} + 1, \, {\rm if} &{\rm sgn} \{\langle  Z_i Z_{i+1} Z_j Z_{j+1} \rangle, \\ & 
\langle  Z_* Z_iZ_{i+1} Z_j \rangle, \langle  Z_* Z_{i} Z_{i+1}  Z_{j+1} \rangle, \langle  Z_* Z_i Z_{j} Z_{j+1} \rangle, \langle  Z_* Z_{i+1} Z_{j} Z_{j+1} \rangle \} \\  & = \{+,-,+,-,+\}\,  {\rm or}\, \{-,+,-,+,-\} \\ 0 & {\rm otherwise} \end{array} \right.
\end{equation}
and we define the total winding number to sum over all the boundaries hit in this way, sign-weighted by the orientation of the boundary in the same way as above: 
\begin{equation}
w_{m=4} = \sum_{i,j} {\rm sgn}(\langle  Z_i Z_{i+1} Z_ j Z_{j+1} \rangle) \times w_{i,j}(Z_*)
\end{equation}
Once again this total winding number is independent of $Z_*$; the argument is exactly the same as we saw above for $m=2$. As we smoothly change $Z_*$, we only have to worry about the situations where a point in the direction of $Z_*$ lies in some two-dimensional boundary of the three-dimensional cell $(i i+1 j j+1)$; i.e. when $x_* Z_* = x_i Z_i + x_{i+1} Z_{i+1} + x_{j} Z_{j+1}$. This boundary is also shared by one other cell $(i i+1 j-1 j)$, and depending on the relative signs of $\langle i i+1 j j+1 \rangle$ and $\langle i i+1 j-1 j \rangle$, either we pass from hitting one boundary to the other as $Z_*$ is smoothly changed with the net contribution to the winding equalling one, or we go from hitting both to missing both with the net contribution being zero. 

This definition of winding generalizes in the obvious way for any even $m$, by counting the number of times a line in the direction $Z_*$ hits the the boundaries $(i_1 i_1+1 \cdots i_{m/2} i_{m/2}+1)$. 

What winding numbers define the amplituhedron? For $m=2$, the winding numbers for $k=1,2,3,4,5,6$ are $1,1,2,2,3,3$, and in general are given by $w_{m=2,k} = \lfloor \frac{(k+1)}{2} \rfloor$. For $m=4$, we have windings $1,1,3,3,6,6$, in general 
\begin{equation}
w_{m=4,k}=\left(\begin{array}{c} \lfloor \frac{k + 3}{2} \rfloor \\ 2 \end{array} \right)
\end{equation}
For general even $m$, the winding number is given by
\begin{equation}
w(m,k) =\left(\begin{array}{c} \lfloor \frac{k + m - 1}{2} \rfloor \\ \frac{m}{2} \end{array} \right)
\end{equation}

We have a simple proof of this fact for the positive Grassmannian case of $n=(k+m)$, and it is empirically correct in all other examples we have checked. It is also interesting to note that $w_{m,k}$ is the {\it maximum} winding possible, so the amplituhedron maximizes winding; we will not prove these statements here, instead giving a simple proof of analogous statements about sign flip patterns in section 5. 

\section{Crossings}

We have seen that for even $m$, the correct topological notion characterizing the amplituhedron is that of ``winding". Already for $m=1$, we have seen that the correct notion was that of counting ``crossings", the number of times the origin $Y$ was crossed in traversals from $1 \to 2 \to \cdots \to n$; this is determined by looking at the number of sign flips in the sequence $\{[\, Y 1 ]\,, \cdots, [\,Y n ]\,\}$. How can we generalize this to general odd $m$? 

In fact the topological notions for odd $m$ and even $(m+1)$ are closely related to each other. Let's consider $m=1,m=2$. In both cases, we look at the collection of simplices $(i i+1)$. Looking at the number of sign flips simply counts how many of these intervals contain the origin (the image of $Y$). In other words, for any interval $(i i+1)$, we define $c_i = +1$ if sgn$\{ \langle i \rangle, \langle i+1 \rangle \} = \{- + \}$ or $\{+ - \}$ and $c_i=0$ otherwise; if $c_i = 1$ the interval $(i i+1)$ contains (or ``crosses") the origin. 

We can extend this idea to any odd $m$. For $m=3$, we look at the exactly the same collection of simplices $(i i+1 j j+1)$ we consider for defining winding for $m=4$. 
Now these 3-dimensional simplices are space-filling in $m=3$ dimensions,  and we can ask how many of them contain the origin. We are then led to define 
\begin{equation}
c_{i,j} =   \left\{\begin{array}{c c}\hspace{0.4cm} + 1, \, {\rm if} &{\rm sgn} \{\langle  Z_{i+1} Z_j Z_{j+1} \rangle,  \langle  Z_{i} Z_j Z_{j+1}\rangle, \langle  Z_i Z_{i+1}  Z_{j+1} \rangle, \langle  Z_i Z_{i+1} Z_{j} \rangle \} \\  & = \{+,-,+,- \}\,  {\rm or}\, \{-,+,-,+\} \\ 0 & {\rm otherwise} \end{array} \right.
\end{equation}

The objects $c_i, c_{i,j}$ are analogous to the $w_i,w_{i,j}$ defined to compute winding numbers. In the winding case, we had to sum over {\it all} the boundaries in order to get an object independent of $Z_*$. For odd $m$, however, the story is a little different. Already for $m=1$ we saw that it was natural to sum over all the boundaries {\it except} the boundary $(n1)$; this already gave us the characterization of the amplituhedron in terms of $k$ sign flips. Of course there would have been no harm in including the $(n1)$ boundary--we would simply add one to the ``crossing" for odd $k$--but it is more natural not to include the $(n1)$ boundary. We will follow this pattern for general odd $m$; we define the crossings to be 
\begin{equation}
c_{m=1} = \sum_{i \neq n} c_i, \,\,\,\,\,\,\, c_{m=3} = \hspace{-0.3cm} \sum_{i<j, j \neq n} c_{i,j}, \, {\rm etc.}
\end{equation}

It is straightforward to compute the number of crossings for $m=3$ by looking at the case of the positive Grassmannian; we find for $k=1,2,3,4,5,6, \cdots$ the crossings $1,2,4,6,9,12$. In general for even $k$ we have $c_k = k(k+2)/4$, while for odd $k$ we have $c_k = ((k+1)/2)^2$, which can be unified in the expression $c_k = \lfloor ((k+1)/2)^2 \rfloor$. 

There is a simple picture relating ``crossing" and ``winding" number that gives us an expression for $c_{k,m}$ for odd $m$ and odd $k$. First, most naively the crossing number for some $k,m$ should naively be double the winding number for $k,m+1$. The reason is that if we start from $m+1$ dimensions and project through some direction $Z_*$, all the boundaries containing the origin will be the ones that were intersected either in the direction $+Z_*$ or $-Z_*$. We can be more precise by 
thinking about about passing from $m+1$ to $m$ dimensions by quotienting through $Z_1 + \epsilon Z_n$.  
Again each winding ``hit" contributes 2 to the crossing number, however, we have to correct for the fact that we ignore the $``1 n"$ facets when counting the crossing number.  But these facets are exactly telling us about what we get for the winding number if we go down into the $(m-1)$ amplituhedron after quotienting by $Z_1$ and $Z_n$.  
Thus for odd $k$, we expect 
\begin{equation}
c_{k,m} = 2 w_{k,m+1} - w_{k,m-1} = \frac{2 k + m - 1}{m+1} \left(\begin{array}{c} \frac{k + m - 2}{2} \\ \frac{m-1}{2} \end{array} \right)\, \, \, ({\rm odd}  \, k)
\end{equation}
On the other hand, for even $m$ we don't get any correction from the $(n1)$ boundaries, and we find 
\begin{equation}
c_{k,m} = 2 w_{k,m+1} = 2 \left(\begin{array}{c} \frac{k + m - 1}{2} \\ \frac{m+1}{2} \end{array} \right) \, \, \, ({\rm even} \, k)
\end{equation}
We have numerically checked the validitiy of these expression up to $m=7$ for large values of $k$. And again, analogous to the statement of maximal winding for even $m$, we have observed that this crossing number is maximized by the amplituhedron.

\section{The Amplituhedron As Binary Code}

The ``winding/crossing" description we have given captures a ``global", topological property of the $Z_a$ data characterizing the amplituhedron. We will now see that this information can even more efficiently be captured in a different way. The key idea is to {\it further} project through some of the external data points, in the only natural way possible, to get down to 1 dimension. It is very easy to see that if we start with some point in a higher $m$ amplituhedron, projecting down to $m=1$ keeps us in the amplituhedron. But remarkably the opposite is also true: the higher $m$ amplituhedron is {\it fully determined} by the requirement that {\it all} possible ``positive projections" down to one dimension land us in the $m=1$ amplituhedron!


Let's begin with the $m=1$ amplituhedron. The claim is that we are in the $m=1$ amplituhedron if and only if the sequence 
\begin{equation}
\{\langle  1 \rangle, \langle  2 \rangle, \cdots, \langle  n \rangle \} \, \textrm { has precisely $k$ sign flips}
\end{equation}
This is equivalent to the characterization of the $m=1$ amplituhedron recently given in \cite{Karp:2016uax}. 

Let's now look at $m=2$. Note that if we project the external ${\cal Z}$ data through ${\cal Z}_1$, the rest of the projected ${\cal Z}$'s are also positive; this is because $\langle {\cal Z}_1 {\cal Z}_{a_1} \cdots {\cal Z}_{a_{k + m - 1}} \rangle > 0$ for $a_1<\cdots<a_{k+m-1}$.  Then it is natural to ask that the projected $Y$ should be in the $m=1$ amplituhedron with external data $({\cal Z}_{2;1}, \cdots, {\cal Z}_{n;1})$ obtained by projecting through ${\cal Z}_1$. We can phrase this purely as a statement about the $m=2$ dimensional data $Z_a$, since projecting through ${\cal Z}_1$ followed by a projection through $Y$ is simply the same as starting from $m=2$ dimensions and projecting through $Z_1$ to get to a one-dimensional space; thus it is natural to ask for the $m=2$ dimensional configuration of the vectors to have the property that when projected through $Z_1$ we land a configuration in the $m=1$ amplituhedron. Now by the twisted cyclic symmetry, we can cycle any one of the ${\cal Z}$'s to the ``${\cal Z}_1"$. Thus, we should demand that {\it no matter which} $Z_a$ we project through, we end up in $m=1$ amplituhedron. Now, we claim that these give us necessary and sufficient conditions for $Y$ to be in the $m=2$ amplituhedron! Said more explicitly, we claim that $Y$ is the $m=2$ amplituhedron if and only if all the following sequences (where $\hat{Z}_i \equiv (-1)^{k-1} Z_i$ accounts for the twisted cyclic symmetry):
\begin{equation}
\left\{\begin{array}{c}\langle  12 \rangle, \cdots \langle 1 n \rangle \\ \langle  23 \rangle  \cdots 
\langle  2 n \rangle \langle  2 \hat{1} \rangle \\ \vdots \\ \langle  n \hat{1} \rangle \cdots \langle  n \widehat{(n-1)} \rangle \end{array} \right\} \, \textrm{ have precisely $k$ sign flips}
\end{equation}
Note as usual that in terms of the underlying $(k+2)$ dimensional data, this is putting constraints on $Y$ since $\langle a b \rangle = [\, Y {\cal Z}_a {\cal Z}_b ]\,$. 

This statement is primary, but we can quickly derive some consequences of it that will lead to a much more efficient check of whether $Y$ is in the $m=2$ amplituhedron. We first observe that the sign-flip conditions trivially reproduce the correct signs of the obvious co-dimension one boundaries of the amplituhedron. For $m=1$, the obvious boundaries are $(-1)^k \langle 1 \rangle >0, \langle  n \rangle > 0$. But this is automatically a consequence of the sequence $\{\langle  1 \rangle, \cdots, \langle  n \rangle\}$ having $k$ sign flips. Now let's look at $m=2$; we will show that the sign flip pattern forces 
\begin{equation}
\langle  i i+1 \rangle > 0 
\end{equation}
Let's start with the sequence 
\begin{equation}
\{\langle  1 2 \rangle, \cdots, \langle  1 n \rangle \}
\end{equation}
Without loss of generality we can set $\langle  1 2 \rangle > 0$. Suppose that $k$ is even; this tells us that $\langle  1 2 \rangle$ and $\langle 1 n \rangle$ are both positive. But now look at the next sequence
\begin{equation}
\{\langle  2 3 \rangle, \cdots, (-1)^{-(k-1)} \langle  2 1 \rangle  \}
\end{equation}
For $k$ even this says that $\langle  2 3 \rangle$ has the same sign as $-\langle  2 1 \rangle = \langle  1 2 \rangle $ and is hence positive.
Continuing in this way we find that all of $\langle  12 \rangle, \langle  2 3 \rangle, \langle  (n-1) n \rangle$ and $\langle  1 n \rangle$ are all positive. The same argument works for $k$ odd. Thus, we see that the sign flip constraint forces the boundaries $\langle  i i+1 \rangle>0$ (where as always $Z_{n+1} = (-1)^{k-1} Z_1$). 

Having established this, we now show that so long as $\langle i i+1 \rangle > 0$, it suffices to check the sign flip pattern for only {\it one} of projections down to $m=1$! In other words, we claim that
\begin{equation}
\begin{array}{c}
Y \textrm{ is in the }  m=2  \textrm{ amplituhedron iff} \\ \, [\, Y i i+1 ]> 0,  \textrm{ and the sequence } \{[\, Y 1 2 ] , \cdots [\,Y 1 n ]\}   \textrm{ has precisely } k \textrm{ sign flips}.
\end{array}
\end{equation}

We now show that all the sign flip patterns follow from just the one beginning with $\langle 1 2 \rangle$ as long as we have $\langle  i i+1 \rangle > 0$. Let's start by showing that if $\{\langle  1 2 \rangle, \cdots \langle  1 n \rangle\}$ has $k$ sign flips, so does 
$\{\langle  2 3 \rangle, \cdots, (-1)^{(k-1)} \langle  2 1 \rangle\}$. 

Let us draw these two sequences one on top of the other, shifted in the natural way: 
\begin{equation} 
\begin{array}{ccccc}
\langle  1 2 \rangle & \langle  1 3 \rangle&  \cdots & \langle  1 n \rangle &  \\
& \langle  2 3 \rangle & \cdots & \langle  2 n \rangle & \langle  2 \hat{1} \rangle  \\
\end{array}
\end{equation}
and let's put in what we already know about the signs:
\begin{equation} 
\begin{array}{ccccc}
 +  & \langle  1 3 \rangle&  \cdots &(-1)^{k-1} &  \\
& + & \cdots & \langle  2 n \rangle & (-1)^{k-1}   \\
\end{array}
\end{equation}
which is clearly compatible with the bottom sequence having $k$ sign flips. Now, since we know what the ends of the sequences look like, let's examine a block of signs in the middle, 
\begin{equation} 
\left\{ \begin{array}{cc} \langle  1 i \rangle & \langle  1 i+1 \rangle \\ \langle  2 i \rangle & \langle  2 i+1 \rangle \end{array} \right\}
\end{equation}
The pattern of these signs cannot be arbitrary. Indeed by the Plucker relation
\begin{equation}
\langle  1 i \rangle \langle  2 i+1 \rangle - \langle  1 i+1 \rangle \langle 2 i \rangle = \langle  i i+1 \rangle \langle  1 2 \rangle > 0 
\end{equation}
where we have used that $\langle 12\rangle$, $\langle  i i+1 \rangle > 0$. Thus while in principle we have $2^4=16$ possible sign patterns in the block, the 4 combinations where $\langle  1 i \rangle \langle  2 i+1 \rangle  < 0$ and $\langle  1 i+1 \rangle \langle 2 i \rangle >0$ cannot occur. The allowed patterns can then be classified as 
\begin{equation}
\left\{ \begin{array}{cc} a & a \\ b & b \end{array} \right\} \, \textrm{ ``don't  change"}, \, \left\{ \begin{array}{cc} a & - a \\ b & - b \end{array} \right\} \, \textrm {``flip both"} \nonumber
\end{equation}
and 
\begin{equation}
\left\{ \begin{array}{cc} a & - a \\ a & a \end{array} \right\} \textrm{ ``flip  top  when  same  as  bottom "}, \, \left\{ \begin{array}{cc} a &  a \\-a & a  \end{array} \right\} \, \textrm{ ``flip  bottom  when  opposite  to  top"} \nonumber
\end{equation}
It is now trivial to see that the number of sign flips in the two sequences must be the same. Obviously the ``don't change" and ``flip both" change the number of flips equally. The crucial point is related to the second set of allowed possibilities. These tell us that if somewhere we have a flip in the top row but not the bottom one, then while we can have any number of flips of both rows thereafter, the {\it next} time there is a flip in one row but not another, it must be that the flip occurs in the second row and not the first! This is because the first row can only flip when it has the same parity as the second, while the second can flip only when it has the opposite parity to the first. 

We can extend this analysis to any higher $m$. Let us illustrate with the case $m=4$. First, if we project the external ${\cal Z}$ data through any $({\cal Z}_b {\cal Z}_{b+1})$, the remaining data will still be positive. So, we claim that $Y$ is the the $m=4$ amplituhedron if and only if, for all such projections, the projected $Y$ is in the $m=2$ amplituhedron; and as we have seen this in turn can be checked by projecting through any $Z_a$ and demanding we end up in the $m=1$ amplituhedron. Thus, more explicitly the claim is that $Y$ is in the $m=4$ amplituhedron iff the sequences (for all $i\neq a,b,b+1$),
\begin{equation}
\{[\, Y a b b+1 i ]\, \} \, \textrm{ have  precisely $k$ sign flips}
\end{equation}
for all $a,b$. As for $m=2$, we can see that this immediately implies that $Y$ is on the right side of the boundaries, i.e. 
\begin{equation}
[\, Y ii+1 jj+1 ]\, > 0
\end{equation}
so that the physics of locality follows from the pattern of sign flips! This follows trivially since we already saw that $\langle  ii+1 \rangle>0$ follows from the sign flip pattern for $m=2$, so if we projected through some $(Z_j Z_{j+1})$ we have $\langle  ii+1 jj+1 \rangle>0$; since we assume the flip pattern must work for all $j$ the result follows. And just as for $m=2$, we will now show that this further implies that we only have to check the sign
flip pattern for a single sequence, that is 
\begin{equation}
\begin{array}{c}
Y \textrm{ is in  the $m=4$ amplituhedron iff} \\ \, [\, Y i i+1 j j+1 ]\, > 0,  \textrm{ and the sequence } \{[\, Y 1 2 3 4 ]\, , \cdots [\,  Y 1 2 3 n ]\, \}  \textrm{ has precisely $k$ sign flips}
\end{array}
\end{equation}
The proof is easy. First, the number of sign flips for the sequences 
$\{\langle 1 j j+1 i \rangle\}$, $\{\langle  2 j j+1 i \rangle\}$, $\{\langle  3 jj+1 i \rangle \},\cdots$ are the obviously the same, since projecting through $(Z_j Z_{j+1})$ we just land on the $m=2$ problem for which we've already established this result. Very slightly more non-trivially we need to show that the number of sign flips for the sequences $\{ \langle  1 j-1 j i \rangle \}$ and $\{\langle  1 j j+1 i \rangle\}$ are the same. But we can easily do this in two steps. First, let's look at the sequences 
$\{\langle  123 i \rangle \}$ and $\{\langle  2 3 4 i \rangle \}$. Since these have $(23)$ in common, projecting through $(Z_2 Z_3)$
lands us on $m=2$ where again we know the number of flips are equal. But then from the fact that the number of flips of $\{\langle  1 jj+1 i \rangle \}$ and $\{\langle  2 jj+1 i \rangle\}$ are the same, we see that the number of flips of $\{\langle  1 2 3 i \rangle\}$ and $\langle 1 3 4 i \rangle \}$ are the same. Continuing in this way we see that the number of flips of $\{\langle  a b b+1 i \rangle\}$ is independent of $a,b$ so long as $\langle  i i+1 j j+1 \rangle > 0$, thus it suffices to only check the sequence $\{\langle 1 2 3 i \rangle \}$. 

For general $m,k,n$, the flip definition of the amplituhedron is then simply the space of $Y$'s for which
\begin{equation}
\begin{array}{c} (-1)^{k} \langle 1 (i_1 i_1 + 1) \cdots (i_{\frac{m-1}{2}} i_{\frac{m-1}{2}+1}) \rangle, 
\langle (i_1 i_1 + 1) \cdots (i_{\frac{m-1}{2}} i_{\frac{m-1}{2}+1}) n\rangle > 0 \textrm{ for $m$ odd} \\ \langle (i_1 i_1 + 1) \cdots (i_{\frac{m}{2}} i_{\frac{m}{2}+1})\rangle
> 0 \textrm{ for $m$ even} \\ \{\langle  12\cdots (m-1) m \rangle,  \cdots, \langle  1 2 \cdots (m-1) n\rangle\}  \textrm{ has $k$ sign flips}
\end{array}
\end{equation}

\subsection*{5.1\,\,\, General Positive Projections and Relations Between Amplituhedra}

Our ``binary code" characterization of the amplituhedron generalizes to  a deeper statement that relates amplituhedra with different values of $m$. To begin with, let us define a ``positive projection" ${\cal P}_{m \to m^\prime}$ to be some $(m-m^\prime)$ plane, such that projecting the ${\cal Z}_a$ data through ${\cal P}_{m \to m^\prime}$ leaves the data positive, that is 
\begin{equation}
[{\cal P}_{m \to m^\prime} {\cal Z}_{a_1} \cdots {\cal Z}_{a_{k+ m^\prime}}]>0 \, {\rm for} \, a_1<\cdots<a_{k+m^\prime}
\end{equation}
Now, it is rather trivial to see, directly from the $Y = C \cdot {\cal Z}$ picture, that if $Y$ is in the amplituhedron for $(m,k)$, then projecting everything through ${\cal P}$, the projected $Y$ is the $(k,m^\prime)$ amplituhedron associated with the projected ${\cal Z}_a$ data. But much more non-trivially, we have an {\it only if} statement: $Y$ is in the amplituhedron if, and only if, for {\it all} positive projections ${\cal P}_{m \to m^\prime}$, the projected $Y$ is the $(k,m^\prime)$ amplituhedron in the projected space. 

The ``binary code" characterization specializes this fact for $m=1$. We also make a somewhat degenerate choice for the positive projections, making use of the fact that if we project through any ${\cal Z}_b,{\cal Z}_{b+1}$, the remaining data is clearly positive. (This is a slightly degenerate choice since ${\cal Z}_b,{\cal Z}_{b+1}$ are projected to the origin). Doing this successively lets us project down to either $m=2$ or $m=1$; further projecting through $Z_1$ also preserves positivity and lets us get from $m=2$ to $m=1$. 

\subsection*{5.2\,\,\, The Positive Grassmannian From Flips}

The case $k=0$ is interesting. Here the ${\cal Z}$ data is simply in the positive Grassmannian of $G_+(m,n)$, and we don't have any $Y$ so that the $Z_a={\cal Z}_a$. It is then interesting to see that our sign flip constraints give a different characterization of the positivity of the $Z$'s. This is trivial for $m=1$; here we say that the sequence $\{\langle 1 \rangle, \langle 2 \rangle , \cdots, \langle n \rangle\}$ has $k=0$ sign flips, which just says that all the entries of the $1 \times n$ $Z$ matrix are positive. We see in general that for $k=0$ we are declaring that certain minors have $k=0$ sign flips, and thus must all have the same sign. Let's now look at $m=2$. Here our criterion is simply that $\langle i i+1 \rangle > 0$, and that $\{\langle 1 2 \rangle, \cdots \langle 1 n \rangle\}$ have zero sign flips; since $\langle 1 2 \rangle >0 $ this just tells us that that the rest of the $\langle 1 i \rangle$ are positive; so for $m=2$ our conditions say that we should have
\begin{equation}
\langle i i+1 \rangle > 0, \, {\rm and} \, \langle 1 3 \rangle, \cdots \langle 1 (n-1) \rangle > 0
\end{equation}
While this doesn't {\it manifestly} force all the ordered minors of $Z$ to be positive, this subset of minors is very well-known to the a ``cluster" of 
$G_{+}(2,n)$; that is, forcing these minors to be positive automatically forces all the rest of the ordered minors to also be positive (on the support of the Plucker relations). 

(We note paranthetically  that here we are taking the ``twisted" cyclic symmetry for granted, but if we back up a step we can actually see its necessity from the sign flip point of view. Suppose we didn't have the twisted cyclic symmetry, but we ask that all the sequences $\{\langle 1 2 \rangle, \cdots \langle 1 n \rangle\}, \{\langle 2 3 \rangle, \cdots, \langle 2 1 \rangle \}$ etc. all have $k=0$ sign flips. Then we quickly run into a contradiction already for $m=2, n=4$: from $\{\langle 1 2 \rangle, \langle 1 3 \rangle, \langle 1 4 \rangle\}$ we would have to say that all these minors are (say) positive, then from $\{\langle 2 3 \rangle, \langle 2 4 \rangle, \langle 2 1 \rangle\}$, since the last sign is negative we would have to say that $\langle 2 3 \rangle, \langle 2 4 \rangle$ are negative, but then finally from $\{\langle 3 4 \rangle, \langle 3 1 \rangle, \langle 3 2 \rangle\}$ we have a contradiction since $\langle 3 2 \rangle$ is forced to be positive while $\langle 3 1 \rangle$ is forced to be negative. So the twisted cyclic symmetry is necessary to get the same number of sign flips through any projections). 
 
The story works the same way for any $m$. Our constraint of $k=0$ sign flips forces a certain set of minors to be positive. For $m$ odd we have that  
\begin{equation}
\begin{array}{c}
\langle 1 (i_1 i_1 + 1) \cdots (i_{\frac{m-1}{2}} i_{\frac{m-1}{2}+1}) \rangle, \langle (i_1 i_1 + 1) \cdots (i_{\frac{m-1}{2}} i_{\frac{m-1}{2}+1}) n\rangle, \\  \, {\rm and} \{\langle 1 2 \cdots (m-1) m \rangle, \cdots \langle 12 \cdots (m-1) n \rangle \} \textrm{ are  all  $>0$} \end{array}
\end{equation}
While for $m$ even these are 
\begin{equation}
\langle (i_1 i_1+1)\cdots(i_{\frac{m}{2}} i_{\frac{m}{2} + 1}) \rangle > 0 \, {\rm and} \, \{\langle 1 2 \cdots (m-1) m \rangle, \cdots \langle 12 \cdots (m-1) n \rangle \} \textrm{ are all $>0$}
\end{equation}
Quite beautifully, the positivity of these minors suffice to force the positivity of all the other minors of $G(m,n)$.  Thus our sign flip criterion successfully (and non-trivially) works for the most trivial case of $k=0$. 

The case where $n=(k+m)$ works in exactly the same way. The external ${\cal Z}$ data can be set to the identity matrix ${\cal Z}_a^I = \delta_a^I$. Let's denote the minors of the $k \times (k+m)$ dimensional $Y$ matrix as $(a_1 \cdots a_k)$. Consider any object of the type $[\, Y b_1 \cdots b_m ]\,$; it is obviously given (up to sign) by the minor $(a_1 \cdots a_k)$, $(a_1 \cdots a_{n-m=k}) = \overline{(b_1,\cdots,b_m)}$ are the conjugate indices to the $(b_1,\cdots, b_m)$. Now, since the sequence $\{[\,Y 1 2 \cdots (m-1) m ]\,, \cdots, [\, Y 1 2 \cdots (m-1) n ]\,\}$ has length $n-(m-1) = k+1$, for this sequence to have $k$ sign flips it must switch signs in every slot, and thus we have sign constraints on the minors of $Y$; of course the boundary constraints also fix signs of the $Y$ minors. For instamce, for $m=2,k=2,n=4$ we have that 
\begin{eqnarray}
[\, Y 1 2 ]\, , [\, Y 23 ]\, , [\, Y 34 ]\, , [\, Y 1 4 ]\, > 0 &\rightarrow& (34), (14),(12),(23) > 0  \nonumber \\ 
{\rm sgn} \, \{[\, Y 1 2 ]\,, [\, Y 1 3 ]\,, [\,Y 1 4 ]\, \} = \{+,-,+\} &\rightarrow& (24)>0
\end{eqnarray} 
and of course the positivity of $(12),(23),(34),(14)$ together with $(24)>0$ also implies $(13)>0$ and so $Y$ is in the positive Grassmannian $G_+(2,4)$.  Conversely, obviously if $Y$ is in $G_+(2,4)$ it will have the correct sign flips. For general $k,m$ with $n=(k+m)$, we force positivity on the ordered minors that are the ``conjugates" 
to the ones we described above for $k=0$, and again there are enough minors to guarantee all the minors are positive. 

\subsection*{5.3\,\,\, The Amplituhedron Maxmizes Flips} 

We pause to note that, so long as the external ${\cal Z}$ data is positive, the {\it maximum} number of flips for our sequences is also given by $k$, in other words, the sequence
\begin{equation}
\{[\, Y 1 2 \cdots (m-1) m ]\, , \cdots, [\, Y 1 2 \cdots (m-1) n ]\, \}\, \textrm{ has {\it at  most}  $k$ sign flips}
\end{equation} The proof uses the same simple observations exploited in the previous subsection. Suppose that there are at least $(k+1)$ sign flips in the sequence, and that they occur at the slots $j_1,j_2, \cdots, j_k,j_{k+1}$, i.e. that we have sgn$([\, Y 123 \cdots (m-1) j_{\alpha +1}]\,) = - ${\rm sgn}$([\,Y 123 \cdots (m-1) j_\alpha ]\,)$. 
Then, ${\cal Z}_1,\cdots,{\cal Z}_{(m-1)},{\cal Z}_{j_1}, \cdots, {\cal Z}_{j_{k+1}}$ are $(m-1) + (k+1) = (k+m)$ vectors that give a basis for the space, so we can expand ${\cal Z}_{j_{k+1}}$ as a linear combination of them; the positivity of the ${\cal Z}$'s fixes the signs in the expansion as described in section 6:
\begin{equation}
{\cal Z}_{j_{k+1}+1} = + {\cal Z}_{j_{k+1}} - {\cal Z}_{j_k} + \cdots + (-1)^k ({\cal Z}_{j_1} - {\cal Z}_{m-1} + \cdots (-1)^{m-1} {\cal Z}_1)
\end{equation}
But then we can compute that $[\, Y 1 2 \cdots (m-1) Z_{j_{k+1}} ]\, = + [\, Y 1 2 \cdots (m-1) Z_{j_k} ]\, - [\, Y 1 2 \cdots (m-1) Z_{j_{k-1}} ]\, + \cdots (-1)^k [\, Y 1 2 \cdots(m-1) j_1 ]\,$; every term on the right-hand side has the same sign as the first term, and so $[\, Y 1 2 \cdots (m-1) j_{k+1}+1]\,$ can't have the opposite sign as  $[\,Y 1 2 \cdots (m-1) j_{{k+1}} ]\,$, contradicting a sign flip at $j_{k+1}$! Thus, a completely equivalent way of characterizing the amplituhedron is simply to say that $Y$ is in the amplituhedron if and only if under any projection to $m=1$ dimensions we have the maximum possible number of $k$ sign flips. 

\subsection*{5.4\,\,\, $Y = C \cdot {\cal Z} \to$ Correct Flips}

We'd like to now show that for for $Y = C \cdot {\cal Z}$ with $C$ in the positive Grassmannian, we have the correct sign-flip pattern. 
First we show that if we've already shown some $C$ gives $Y = C \cdot {\cal Z}$ with the correct flips, then we can always add zero columns to $C$ without changing the conclusions. The argument is trivial for even $m$, since we can always use the cyclic symmetry to put the zero column at the very end. But then we are merely adding a last $[\, Y 1 \cdots (m-1) (n+1) ]\,$ to our sequence, and since we already have the maximum number $k$ of flips we can't have any more. In this way, by adding a zero column at the end and then cyclically shifting, we can add zeroes in any columns we like without changing the total number of sign flips. Since we've already proven than $Y= C \cdot {\cal Z}$ for the $n=(k+m)$ case, for $n>(k+m)$, we have also established the right flip pattern for the image of those $k \times n$ dimensional cells of $G_+(k,n)$ which correspond to positive matrices in a $(k+m)$ subset of the $n$ columns. But we would like to show that for ${\it any}$ positive matrix $C_{\alpha a}$, the projection through $Y = C \cdot {\cal Z}$ gives the right sign flip pattern. 

Here we make use of a simple but non-trivial fact about positive matrices, which tells us how to systematically build more complicated positive matrices from simpler ones. Any $K \times N$ matrix in the positive Grassmannian, including generic points in the interior (or the ``top cell"), can be constructed starting from some zero-dimensional cell (corresponding to the $(K \times N)$ matrix being set to the identity in some $(K \times K)$ block and vanishing elsewhere), and recursively shifting the columns of the matrix by positive multiples of its immediately neighboring (non-vanishing) columns. 

Thus, we can make any positive matrix $C$, by beginning with zero dimensional cells where the $C$ matrix is the identity in some $k \times k$ block, and then repeatedly shifting a given column of $C$ by positive multiples of its neighbors. But note that under $C_{\alpha a} \to C_{\alpha a} + x_{a+1} C_{\alpha a+1}$, the effect on $Y$ is the same as if we shifted ${\cal Z}_a \to {\cal Z}_a + x_{a+1} {\cal Z}_{a+1}$; since this preserves the positivity of the ${\cal Z}$, again the (maximized) number of flips can not be altered. In this way we can work our way up from $C$'s corresponding to zero-dimesional cells of $G_+(k,n)$ to any point in $G_+(k,n)$. 

The only  subtlety in this argument is that at the starting point, where $C$ is a zero-dimensional cell fixed to the identity matrix in columns $(i_1, \cdots, i_k)$, $Y=({\cal Z}_{i_1} \cdots {\cal Z}_{i_k})$ is also on a zero-dimensional boundary of the amplituhedron, and many of the brackets $[\, Y 1 2 \cdots (m-1) i ]\,$ vanish and so there is ambiguity in how to assign the signs and decide whether the starting flip pattern in correct. But there is a very easy fix to this problem. We simply choose $C$ to be in the positive Grassmannian associated only with columns $(i_1, \cdots, i_k)$ and any $m$ other columns, with tiny values for positive co-ordinates chosen so that $C$ is very close to the zero-dimensional cell which is the identity in $(i_1,\cdots,i_k)$. Since we have already established that we get the correct sign-flip pattern for this case, we have done what was needed---find a slight deformation that has the correct sign flip pattern. Starting from this point, we do exactly the shifts of columns of $C$ by adjacent columns that takes $C$ to a generic point in $G_+(k,n)$ and the argument follows as before; the number of flips is preserved in every step and we etablish the claimed result.

\section{Factorization}

One of the central features of amplituhedron geometry is the way in which the co-dimension one boundaries of the amplituhedron are closely related to amplituhedra with lower $k$ and $n$. This is expected to be a feature of amplituhedra for all $m$. In the particular case of $m=4$ we expect to see the the amplituhedron with some $k,n$ ``factorize" into two lower-point amplituhedra $(k_L,n_L)$ and $(k_R,n_R)$ with $n_L + n_R = n+2$ and $k_L + k_R = k-1$. We can see an avatar of factorization in ``$C \cdot {\cal Z}$" description of the amplituhedron, in the form of the $C$-matrices on co-dimension one boundaries of the space. For instance when $m=4$, on the co-dimension-one boundaries where $[Y ii+1 jj+1] \to 0$, we can write $Y = y_f y$ where  $y_f$ is a point in the span of $({\cal Z}_i, {\cal Z}_{i+1},{\cal Z}_j, {\cal Z}_{j+1})$ and $y$ is a $(k-1)$ plane. This implies that the $C$ matrix should have a representation where the top row is non-zero only in the entires $(i i+1 j j+1)$. But then, remarkably, positivity forces $C$ to ``factorize" in the form 

$$
\includegraphics[scale=.55]{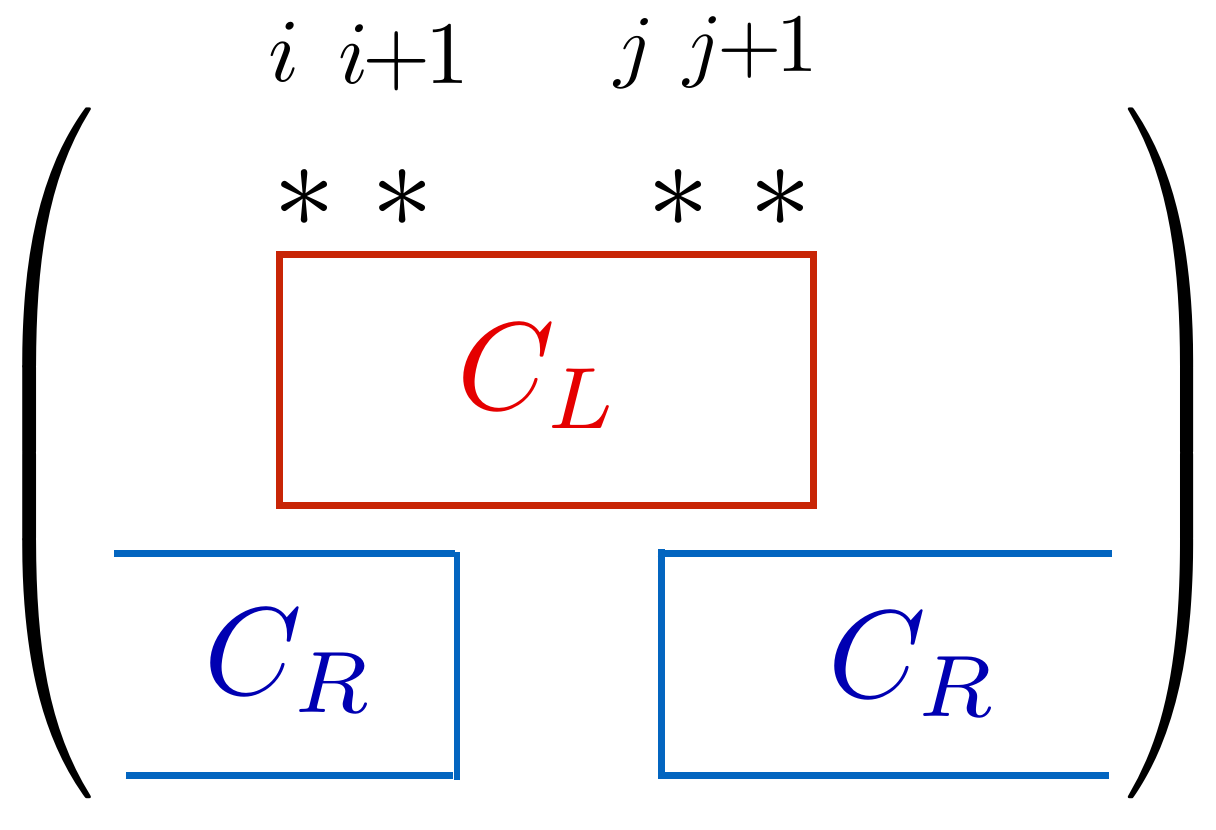}
$$
where the blocks $C_L$ and $C_R$ are individually positive. This is strongly suggestive of factorization for the amplituhedron geometry itself. Let us examine this geometry more precisely. Given the point $y_f$ in $Y$ which is in the span of $({\cal Z}_i, {\cal Z}_{i+1},{\cal Z}_j, {\cal Z}_{j+1})$,  we can expand $y_f = (\alpha_i {\cal Z}_i + \alpha_{i+1} {\cal Z}_{i+1}) + (\alpha_{j} {\cal Z}_j + \alpha_{j+1} {\cal Z}_{j+1}) \equiv I_i + I_j$. 
Then if we project through $y_f$, the geometry should consist of ``left" and ``right" amplituhedron, where the external data of the ``left" are (the projections through $y_f$ of) $I_i, {\cal Z}_{j+1}, \cdots, {\cal Z}_i$ and the ``right" amplituhedron has external data ${\cal Z}_{i+1}, \cdots, {\cal Z}_j, I_j$. (It is easy to see that this projected data is positive). 
While this fact is strongly suggested by the ``factorization" of the $C$-matrix, it is not easy to prove from the $C \cdot {\cal Z}$ picture; for instance it is not obvious that the different $(k_L,n_L);(k_R,n_R)$ splits are all non-overlapping in $Y$ space. 
As we will now see, the factorization structure of the amplituhedron boundary follows simply and provably from our point of view, as an elementary consequence of the ``binary code" of sign-flip patterns. 

Let's start with $m=2$. The factorization picture we expect is the following. The boundaries are at $[\, Y i i+1 ]\, \to 0$; without loss of generality we will consider the boundary where $[\, Y 1 2 ]\, \to 0$. We can set $Y = y_f y$ where $y_f = ({\cal {\cal Z}}_1 + x {\cal {\cal Z}}_2)$ with $x>0$, and $y$ is a $(k-1)$ plane. If we project through $y_f$, the resulting projected data $({\cal {\cal Z}}_{2,f} \cdots,{\cal {\cal Z}}_{n,f})$ is still positive. The ``factorization" statement is then that $y$ is in the $k-1$ amplituhedron. Said in terms of sign flips, this means that as we take $[\, Y 1 2 ]\, \to 0$,  the sequence $\{[\, Y  2 3 ]\,, [\, Y 2 4 ]\,, \cdots, [\, Y 2 n ]\,\}$ has $(k-1)$ flips. 

The heart of the matter will be to show that if $[\, Y 1 2 ]\, \to 0$, then necessarily $[\, Y  1 3 ]\,< 0$. Let us assume for this for the moment and show how our desired result follows from it. Let's write again $Y = ({\cal {\cal Z}}_1 + x {\cal {\cal Z}}_2) y$, then if $[\, Y 1 3 ]\, < 0$ we have that $x[\,2 y 1 3 ]\, < 0$, but we also know that $[\,Y 2 3 ]\,> 0$ which means that $[\, 1 y 2 3 ]\,> 0$; thus we must have $x>0$. Now we are interested in the sign pattern of the sequence $\{{\rm sgn}([\, Y 2 i ]\,)\} = \{{\rm sgn}([\,1 y 2 i ]\,)\}$. But this can clearly related to the sign pattern of the sequence $\{{\rm sgn}([\, Y 1 i ]\,) \} = \{{\rm sgn}( x [\,2 y 1 i ]\,)\} = - \{{\rm sgn}([\, Y 2 i ]\,\}$. Thus, the number of sign flips of the sequence $\{[\,Y 23 ]\,, \cdots, [\, Y 2 n ]\, \}$ is the same as counting the number of sign flips of $\{[\, Y 1 3 ]\,, \cdots [\, Y 1 n  \}$.

Now we know that the sequence $\{[\, Y 1 2 ]\,, [\, Y 1 3 ]\,, \cdots, [\, Y 1 n ]\,\}$ has $k$ sign flips; even though on the boundary we have $[\, Y 1 2 ]\, \to 0$, it was approached from $[\, Y 1 2 ]\, >0$. Furthermore since $[\, Y 1 3 ]\, < 0$, we started this sequence with a single flip. Therefore, the rest of the sequence $\{[\, Y 1 3 ]\,, \cdots [\, Y 1 n ]\,\}$ must have $(k-1)$ flips, as desired. 

So we now simply have to prove that as $[\, Y 1 2 ]\, \to 0$, we must have $[\, Y 1 3 ]\, < 0$. The proof will importantly use both the fact that the sequence $\{[\, Y 1 i ]\,\}$ has $k$ sign flips, as well as sign patterns associated with the positivity of the 
${\cal Z}$ data. 

Suppose to the contrary that $[\, Y 1 3 ]\, > 0$. Then we must have $k$ places to the right of $3$ where where the sign flips occurred, let's call then $b_1, \cdots, b_k$; in other words we must have the signs
\begin{equation}
\left\{ \begin{array}{cccccc} [\, Y 1 2 ]\, &  [\, Y 1 3 ]\, & [\, Y 1 b_1 ]\, & [\, Y 1 b_2 ]\, & \cdots & [\, Y 1 b_k ]\, \\
0^+ & + & - & + & \cdots & (-1)^k \end{array}\right\}
\end{equation}
We will now expand ${\cal Z}_{b_1}$ in terms of the basis of ${\cal Z}_1,{\cal Z}_2,{\cal Z}_3,{\cal Z}_{b_{2}}, \cdots,{\cal Z}_{b_{k}}$, and here the positivity of the ${\cal Z}$ data will be  important, since it implies a fixed pattern of signs in this expansion. 

Indeed let us consider more generally  $n$ vectors ${\cal Z}_a$ in $K$ dimensions, with all ordered minors positive. 
Let us consider $(K+1)$ of these vectors. Then we can expand any one of them in a basis of the other $K$; the positivity of the ordered minors implies certain sign patterns on the coefficients of this expansion. For instance consider $K=4$ and any five ${\cal Z}_{a_1}, \cdots, {\cal Z}_{a_5}$ for $a_1<a_2<\cdots<a_5$. Then, we can for instance expand ${\cal Z}_{a_1}$ in a basis of the rest: 
\begin{eqnarray}
{\cal Z}_{a_1}& = &
\frac{[a_1 a_3 a_4 a_5]  {\cal Z}_{a_2} - [ a_1 a_2 a_4 a_5 ] {\cal  Z}_{a_3} + [ a_1 a_2 a_3 a_5 ] {\cal Z}_{a_4} - [a_1 a_2 a_3 a_4 ] {\cal Z}_{a_5}}{[ a_2 a_3 a_4 a_5 ]} \nonumber \\ &=& +  {\cal Z}_{a_2} - {\cal Z}_{a_3} + {\cal Z}_{a_4} - {\cal Z}_{a_5}
\end{eqnarray}
where in the second expression we are only keeping track of the signs of the coefficients. More generally, for positive $Z$ and any ordered $a_1< \cdots<a_{K+1}$, any given $Z_{a_l}$ can be expanded in terms of the others, starting with $+$ signs for its immediate neighbors to the left and right and alternating signs both to the left and to the right: 
\begin{equation}
{\cal Z}_{a_l} = \begin{array}{c} + {\cal Z}_{a_{l+1}} - {\cal Z}_{a_{l+2}} + \cdots \\ + {\cal Z}_{a_{l-1}} - {\cal Z}_{a_{l-2}} + \cdots \end{array}
\end{equation}

Applying this general fact to our case of interest we have simply
\begin{equation}
{\cal Z}_{b_1} = \begin{array}{c} + {\cal Z}_3 - {\cal Z}_2 + {\cal Z}_1 \\ + {\cal Z}_{b_2} - {\cal Z}_{b_3} + \cdots + (-1)^k {\cal Z}_{b_k} \end{array}
\end{equation}
But using this expansion we can compute
\begin{equation}
\begin{array}{c} [\, Y 1 b_1 ]\, = [\, Y 1 3 ]\, - [\, Y 1 2 ]\, + [\, Y 1 b_2 ]\, - [\, Y 1 b_3 ]\, + \cdots + (-1)^k [\, Y 1 b_k ]\, \\ = (+) + (0) + (+ ) + (+) + \cdots + (+) > 0 \end{array}
\end{equation}
which contradicts $[\, Y 1 b_1 ]\,  < 0$. Thus we can't have $[\, Y 1 3 ]\,>0$, and must have $[\, Y 1 3 ]\, < 0$. 

Let us now move on to the more interesting case $m=4$. Suppose we are sitting on the boundary where $[\, Y 12 jj+1 ]\, \to 0$. By projecting through either $(12)$ or $(jj+1)$ to get to $m=2$, we can conclude that $Y = ({\cal Z}_1 + x {\cal Z}_2 + x_j {\cal Z}_j + x_{j+1} {\cal Z}_{j+1}) y$ where $y$ is a $(k-1)$ plane and $x>0$, with $x_j,x_{j+1}$ having the same sign. Also, from what we've just learned about $m=2$, projecting through $(jj+1)$ we can conclude that the sequence 
\begin{equation}
\{[\, Y 2 3 j j+1 ]\,, [\, Y 2 4 j j+1 ]\,, \cdots, [\, Y 2 (j-1) j j+1 ]\,; [\, Y 2 (j+2) jj+1 ]\,, \cdots, [\, Y 2 n j j+1 ]\,\}
\end{equation}
has $(k-1)$ sign flips. But from the facts that $[\, Y 1 2 j-1 j ]\, > 0, [\, Y 1 2 j j+1 ]\, > 0$, we conclude that $x_{j+1} [\, y 1 2 j-1 j j+1 ]\,>0$ and $x_{j} [\, y 1 2  j j+1 j+2 ]\, > 0$; since $x_j,x_{j+1}$ have the same sign we conclude that $[\, y 12 j-1 j j+1]$ and $[\, y 1 2 j j+1 j+2 ]\,$ have the same sign. But this means that $[\, Y 2 (j-1) j j+1]\, = (-1)^{k-1} [\, y 1 2 (j-1) j j+1 ]\,$ and $[\, Y 2 (j+2 ) j j+1 ]\, = (-1)^{k-1} [\, y 1 2 j j+1 j+2]\,$ have the same sign. Given the above sequence has $(k-1)$ sign flips and since we have seen that $[\, Y 2 (j-1) j j+1 ]\,; [\, Y 2 (j+2) jj+1 ]\,$  have the same sign, there is no sign flip at those slots, so we conclude that 
\begin{equation}
\begin{array}{cc} \{[\, Y 2 3 jj+1 ]\,, \cdots, [\, Y 2 (j-1) j j+1 ]\,\} & \textrm{ has $k_R$ sign flips} \\ \{[\, Y 2 (j+2) j j+1 ]\,, \cdots, [\, Y 2 n j j+1 ]\,\} & \textrm{ has  $k_L$ sign flips} \end{array},  {\rm with}\, k_L + k_R = k - 1
\end{equation}
Now note that since $(jj+1) = (j I_j)$, the first sign sequence above is precisely what we would look at to check membership in the $k_L$ amplituhedron with external data ${\cal Z}_2, \cdots, {\cal Z}_j, {\cal Z}_{I_j}$. Note also that $[\, Y 1 2 (j+1) i ]\, =x_j [\, j y 1 2 j+1 i ]\, = -x_j [\, Y 2 i j j+1 ]\,$. Thus the number of sign flips of the second sequence is exactly the same as the sequence $\{[\, Y 1 2 (j+1) (j+2), \cdots, [\, Y 12 (j+1) n ]\,\}$; since $12 = 1 I_1$, this precisely checks membership in the $k_L$ amplituhedron with external data $({\cal Z}_1,{\cal Z}_{I_1},{\cal Z}_{(j+1)}, \cdots, {\cal Z}_n)$. 

Strictly speaking, this argument tells us that every point on the boundary of the amplituhedron belongs to the factorized product of the lower amplituhedra, but the possibility is left open that the amplituhedron boundary is only a subset of the sum of the product of lower amplituhedra and does not fully cover it. However, since we have shown that all $Y=C \cdot {\cal Z}$ do have the right flip count, we know that all the image of all the $C$ matrices of the factorized form will have the correct flip counts on both the left and right, and we are done. 

\section{Triangulations from Sign Flips}

For $m=1$ and $m=2$, keeping track of the sign flip pattern give us a natural triangulation of the amplituhedron. Let's consider first $m=1$, where $Y$ is a $k$-plane in $(k+1)$ dimensions. Start with the easiest case $k=1$. Since we know $\{[\, Y 1 ]\,, \cdots, [\, Y n ]\,\}$ has one sign flip, let's focus on the place this flip takes place; there is some $j$ for which $[\, Y j ]\, <  0$ but $[\, Y (j+1) > 0$. The full $m=1,k=1$ amplituhedron is then covered for the collection of these regions for all $m$. Now, with $k=1$ we can always expand $Y$ in some basis ${\cal Z}_A,{\cal Z}_B$ as $Y$ as $Y = {\cal Z}_A + x_B  {\cal Z}_B$; in order to describe the $m=1$ ``cell" where the sign flip occurs in the $j$'th slot, it is clearly convenient to choose ${\cal Z}_A = {\cal Z}_j$ and ${\cal Z}_B = {\cal Z}_{j+1}$. Then we see that $[\, Y j ]\, = - x [\, j j+1 ]\,, [\, Y j+1 ]\, = [\, j j+1 ]\,$. Thus to match the sign pattern in this cell we must have $x>0$; and conversely, every $Y$ of this form with $x>0$ will belong to this cell. We can proceed in the same way to $k=2$. Here we can characterize {\cal Z} the sign flips completely by specifying the two slots in the flips took place; so there is some $j_1$ and $j_2$ for which $[\, Y j_1 ]\, > 0, [\, Y (j_1 + 1) ]\, < 0, [\, Y j_2 ]\, < 0, [\, Y (j_2+1) ]\, > 0$. Again we can conveniently expand $Y = ({\cal Z}_{j_1} + x_1 {\cal Z}_{j_1 + 1})({\cal Z}_{j_2} + x_2 {\cal Z}_{j_2 + 1})$. 
Now $[\, Y (j_1 + 1) ]\, < 0$ tells us that $[\, {\cal Z}_{j_1} {\cal Z}_{j_1 + 1} ({\cal Z}_{j_2} + x_2 {\cal Z}_{j+2}) ]\, > 0$, so then the positivity of $[\, Y j_1 ]\,  = x_1  [\, {\cal Z}_{j_1} {\cal Z}_{j_1 + 1} ({\cal Z}_{j_2} + x_2 {\cal Z}_{j_2 + 1})]\,$ tells us we must have $x_1 > 0$. Similarly $x_2>0$. And again conversely, every $Y$ of the form with $x_1,x_2>0$ will belong to this ``cell" of the $m=1,k=2$ amplituhedron. In general then, we find that 
\begin{equation}
\begin{array}{c} \textrm{ The region in the $m=1$ amplituhedron where }  \{[\, Y 1 ]\, \cdots [\, Y n ]\,\} \, \textrm{ flips in slots } j_1, \cdots j_k \\ \textrm{ is  covered  by } \\ Y =({\cal Z}_{j_1} + x_1 {\cal Z}_{j_1 + 1})({\cal Z}_{j_2} + x_2 {\cal Z}_{j_2+1}) \cdots({\cal Z}_{j_k} + x_k {\cal Z}_{j_k + 1}) \, \textrm{ with } x_k \geq 0 \end{array}
\end{equation}
We can trivially relate this to the $``Y = C.{\cal Z}"$ description of the amplituhedron; we can think of $Y$ as the span of the $k$ points of $Y_\alpha$ with $Y_\alpha = {\cal Z}_{j_\alpha} + x_\alpha {\cal Z}_{j_\alpha + 1}$. Then we can also recognize this as 
$Y_\alpha = C_{\alpha a} {\cal Z}_a$,  where  
\begin{equation}
C^{\{i_1,\cdots,i_k\}}_{\alpha a} = \left\{\begin{array}{cc} 1 &  a = i_{\alpha} \\  x_\alpha &  a=i_{\alpha} + 1 \\ 0 & {\rm otherwise} \end{array} \right\}
\end{equation}
with the positive variables ${\cal Z}_\alpha \geq 0$. Note that the ordered minors of this $C$-matrix are all positive. 

The $k$ form associated with this cell is 
\begin{equation}
\Omega^{\{i_1, \cdots, i_k\}} = \prod_{\alpha=1}^k d {\rm log}\,x_\alpha = \prod_{\alpha=1}^k d {\rm log}\left(\frac{[\, Y {\cal Z}_{i_\alpha+1} ]\,}{[\, Y {\cal Z}_{i_\alpha} ]\,}\right) 
\end{equation}
and the full form is 
\begin{equation}
\Omega = \sum_{1 \leq i_1 < \cdots i_k \leq (n-1)} \Omega^{\{i_1, \cdots, i_k\}}
\end{equation}

The $m=2$ amplituhedron can be triangulated in precisely the same way. The only difference is that we have to mark the slots $(j_1,\cdots,j_k)$ 
where the sequence $\{[\, Y 1 2 ]\,, \cdots, [\, Y 1 n ]\,\}$ has its sign flips. Now $Y$ is a $k$-plane in $(k+2)$ dimensions, and we can parametrize any $k$-plane as $Y=(+{\cal Z}_1 + x_{1} {\cal Z}_{j_1} + y_{1} {\cal Z}_{j_1 + 1})(-{\cal Z}_1 + x_{2} {\cal Z}_{j_2} + y_{2} {\cal Z}_{j_2 + 1})(+{\cal Z}_1 + x_3 {\cal Z}_{j_3} + y_3 {\cal Z}_{j_3 + 1})\cdots((-1)^k {\cal Z}_1 + x_k {\cal Z}_{j_k} + y_k {\cal Z}_{j_k + 1})$; here the alternating signs in front of ${\cal Z}_1$ are chosen for convenience. Then just as for $m=1$, the pattern of signs forced by flips at $j_1,\cdots,j_k$ forces all the $x_\alpha, y_\alpha \geq 0$, and conversely any $Y$ of this form has flips in these slots. We can think of these ``cells" in the $Y = C\cdot{\cal Z}$ language as $Y_\alpha = (-1)^{(\alpha-1)} {\cal Z}_1 + x_\alpha {\cal Z}_{j_\alpha} + y_\alpha {\cal Z}_{j_\alpha + 1}$, giving us a $C$ matrix of the form 
\begin{equation}
C^{\{j_1,\cdots,j_k\}}_{\alpha a} = \left\{\begin{array}{cc} (-1)^{\alpha -1 }  & a =1 \\ x_\alpha  & a=j_{\alpha} \\ y_{\alpha} & a=j_\alpha + 1 \\ 0  & {\rm otherwise} \end{array} \right\}
\end{equation}
with manifestly positive minors. 

We can also see this triangulation very naturally from the winding picture. Let us characterize the winding pattern by looking at the boundaries that are hit when we choose ${\cal Z}_*$ to point in the two directions ${\cal Z}_* = -{\cal Z}_1$, and ${\cal Z}_* = +{\cal Z}_1$; equivalently we are looking at which boundaries are intersected by the full line joining ${\cal Z}_1$ and the origin. For $k=1$, since we have winding number 1 the line in the direction $-{\cal Z}_1$ must intersect a single boundary $(i_1 i_1+1)$. The direction $+{\cal Z}_1$ is degenerate since both $(n1)$ are $(12)$ are hit; a small variation means only one of the two is hit. So it is useful to characterize a cell just by the $(i_1 i_1+1)$ boundary hit in the direction $-{\cal Z}_1$. Next let's look at $k=2$. The winding number here is again 1, and the direction $-{\cal Z}_1$ again hits some $(i_1 i_1+1)$. But now in the direction $+{\cal Z}_1$, we find that a small variation will either cause the line to hit both of $(n1),(12)$ or miss both of them. Thus winding number 1 means that in the direction $+{\cal Z}_1$ some other boundary $(i_2 i_2 + 1)$ is hit. Then for $k=3$ with winding number 2, in direction $-{\cal Z}_1$ we must hit two boundaries $(i_1 i_1+1),(i_2 i_2+1)$, while in (a small deformation of) the direction $+{\cal Z}_1$ we hit one of $(n1),(12)$, and thus one more boundary $(i_3 i_3+1)$ is hit. This pattern obviously continues for all $k$: the line joining ${\cal Z}_1$ to the origin intersects $k$ boundaries $(i_1 i_1+1), \cdots, (i_k i_k+1)$. This picture corresponds precisely to what we would see by projecting through ${\cal Z}_1$,  and the cells correspond to exactly the same one we arrived at from the flip picture. 

$$
\includegraphics[scale=.6]{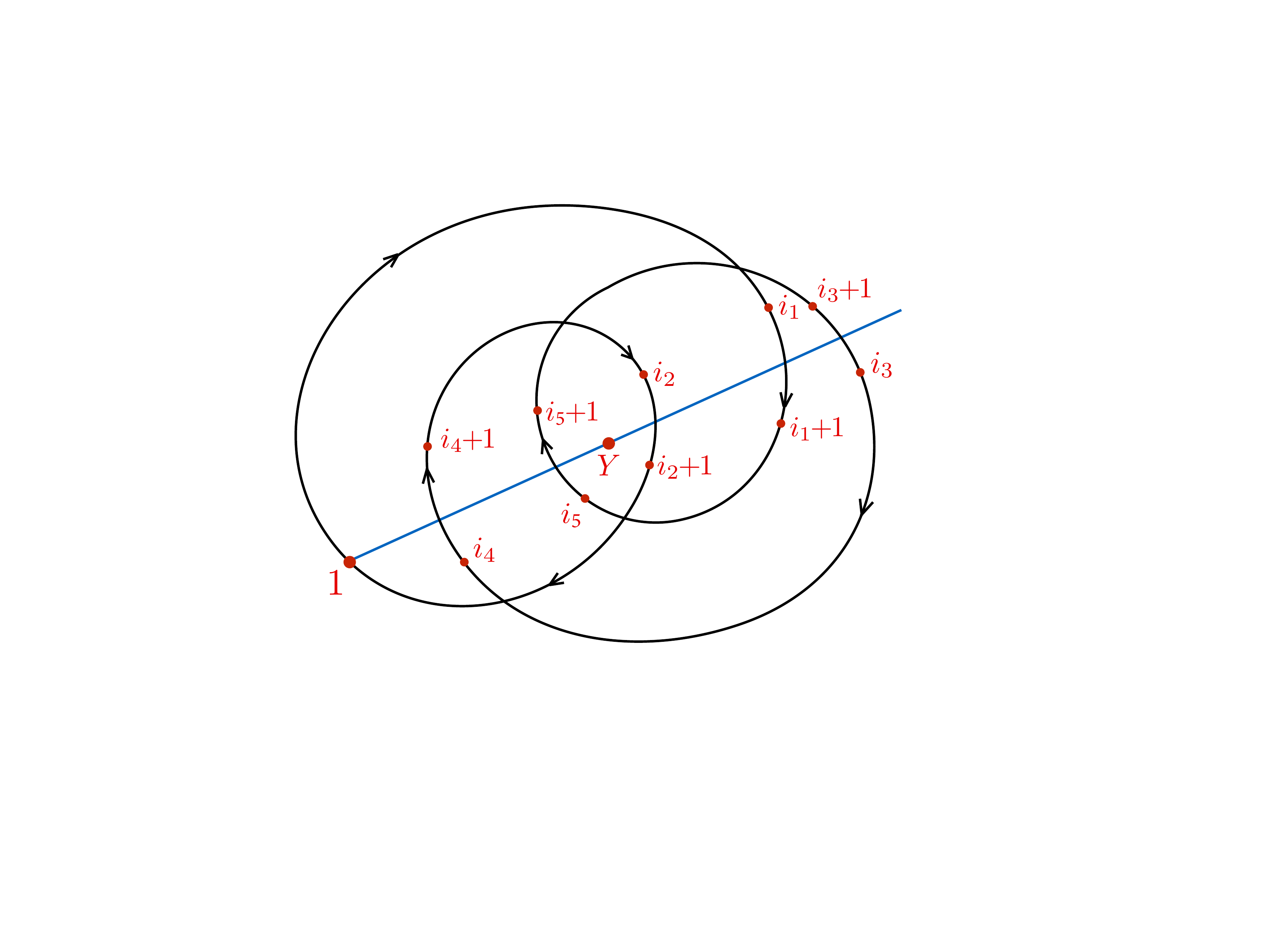}
$$

The $2k$ form associated with this cell is 
\begin{align}
\Omega^{\{i_1, \cdots, i_k\}} &= \prod_{\alpha} d {\rm log}\,{x}_\alpha\,\, d{\rm log}\,y_\alpha \\ 
&\hspace{-1cm}= \prod_{\alpha} d {\rm log} \left(\frac{[\, Y {\cal Z}_1 {\cal Z}_{i_\alpha} ]\,}{[\, Y {\cal Z}_{i_\alpha} {\cal Z}_{{i_\alpha}+1} ]\,}\right) d {\rm log} \left(\frac{[\, Y {\cal Z}_1 {\cal Z}_{i_\alpha+1} ]\,}{[\, Y {\cal Z}_{i_\alpha} {\cal Z}_{{i_\alpha}+1}} ]\, \right) \nonumber \\ 
&\hspace{-1cm}=\frac{{\rm d}^{k(k+2)}Y}{{\rm Vol(GL}(k))} \frac{\left[[\, \left(Y^{k-1}\right)^{\alpha_1} {\cal Z}_1 {\cal Z}_{i_1} {\cal Z}_{i_1+1}]\, [\, \left(Y^{k-1}\right)^{\alpha_2} {\cal Z}_1 {\cal Z}_{i_2} {\cal Z}_{i_2+1}]\, \cdots [\, \left((Y^{k-1}\right)^{\alpha_k} {\cal Z}_1 {\cal Z}_{i_k} {\cal Z}_{i_k+1}]\,  \epsilon_{\alpha_1 \cdots \alpha_k}\right]^k}{\prod_\alpha [\, Y {\cal Z}_1 {\cal Z}_{i_\alpha} ]\, [\, Y {\cal Z}_1 {\cal Z}_{i_{\alpha}+1} ]\, [\, Y {\cal Z}_{i_\alpha} {\cal Z}_{i_\alpha+1} ]\,} \nonumber
\end{align}
As usual the full form arise from summing over the form for each piece of the triangulation 
\begin{equation}
\Omega = \sum_{1 \leq i_1 < \cdots i_k \leq (n-1)} \Omega^{\{i_1, \cdots, i_k\}}
\end{equation} 
This form also has spurious poles that cancel between the terms.

This most direct connection between sign patterns and triangulations of the amplituhedron is restricted to the simplest $m=1,2$ amplituhedra. Starting with $m=3,4$, there isn't a simple relation between the image of a particular cell of $G_+(k,n)$, and any one sign pattern.

\section{Loops}

We now move on to loops, beginning with a quick review of the usual definition of the loop-level amplituhedron. We fix $m=4$; at $L$ loops, we have $(k+2)$-planes $(YAB)_i$ for $i=1, \cdots, L$, all of which intersect on a common $k$-plane $Y$. 
We can describe any of the planes $(YAB)_i$ as the span of $Y$ together with a 2-plane $(AB)$; together with a redundancy that allows us to translate $(AB)$ in any direction of $Y$. If we denote $(AB)_i$ by $(A_1 A_2)_i$, the $L$-loop amplituhedron is defined to be all the $Y_\alpha^I$ and $A^I_{\sigma; i}$ of the form 
\begin{equation}
Y_\alpha^I = C_{\alpha a} {\cal Z}_a^I, \, A_{\sigma, i} = D^{(i)}_{\sigma a} {\cal Z}_a^I
\end{equation}
where we have new $(2 \times n)$ matrices $D^{(i)}_{\sigma, a}$ which are defined up to translations by the $C_{\alpha a}$. Together with $C_{\alpha a}$, these satisfy an extended positivity constraint of the ``loop-positive Grassmannian", which say that, for any collection of $0 \leq l \leq L$ of the $D$'s, $D^{(i_1)}, \cdots, D^{(i_l)}$, the ordered minors of the matrix 
\begin{equation}
\left(\begin{array}{c} D^{(i_1)} \\ \vdots \\ D^{(i_l)} \\ C \end{array} \right)
\end{equation}
are all positive. 
 
Let us turn to extending our topological characterization of the the amplituhedron to loop level. When we project through $Y$, $YAB$ projects down to a 2-plane we can call $(AB)$. 
It is then natural to conjecture the following: projecting through $(YAB)$ the 2-dimensional data should correspond to the $m=2, (k+2)$ amplituhedron, while projecting through $Y$ we should end up in the $m=4,k$ amplituhedron as usual. 
As before, we can phrase this in terms of projections down to the $m=1$ amplituhedron, which tells us that for fixed $a$,$b$, 
\begin{equation}
\{ [(Y A B) a i ]\} \, \textrm{ has } (k+2)  \textrm{ sign flips for all } a, \, \{[Y a b b+1 i ]\}\, \textrm{ has } k  \textrm{ sign  flips  for  all} \, a,b
\end{equation}
Again as before, this has the effect of requiring that $[(Y A B) i i+1]>0,  [Y ii+1 j j+1]>0$,
and if we assume these conditions, then it suffices to check the sign flip pattern only through one set of projections. This leads to the most efficient characterization of the 1-loop amplituhedron as those $(YAB)$, $Y$ for which
\begin{eqnarray} 
[(Y A B) i i+1]> 0, [Y ii+1 jj+1]>0 \\ 
\{[(Y A B) 1 2],\cdots,[(Y A B) 1 n ]\} \, \textrm{ has } (k+2) \textrm{ sign flips} \\
\{[Y 1 2 3 4], \cdots, [Y 1 2 3 n]\} \, \textrm{ has }  k \textrm{ sign flips} 
\end{eqnarray}
When there is more than one loop, we have several $(k+2)$-planes $(YAB)_\gamma$, with the $k$-plane $Y$ common to all of them. The conditions are exactly the same as the above for each loop separately. But it is also natural to demand after projecting through any of the $(AB)_\gamma$ to get to a 2-dimensional space, that further projecting through $(AB)_\rho$ should land us in the ``$m=0$ amplituhedron", which is just the condition that $[Y (AB)_\gamma (AB)_\rho]>0$. 

Note that this definition gives us an extremely simple picture for the loop amplituhedron. At one loop, we simply have that the amplituhedron is the intersection of the $m=2, (k+2)$ and the $m=4,k$ ``tree" amplituhedra! That is, $(YAB,Y)$ is in the 1-loop amplituhedron, if the $(k+2)$ plane $(YAB)$ is in the $m=2$ amplituhedron, with the $k$-plane $Y$ inside $(YAB)$ is in the $m=4$ amplituhedron.  And for any number of loops we have the further intersection with the ``$m=0$" amplituhedron. 
None of this is obvious from the $``(C,D)"$ picture of the loop amplituhedron, and this even suggests new approaches to triangulating the amplituhedron. 

Consider for example the case of $k=1,L=1,n=5$. In the new picture, we simply have $(YAB)$ in the $m=2,n=5,k=3$ amplituhedron where it is just the $G_+(3,5)$ positive Grassmannian. Now this plane slices through the tree amplituhedron (just the polytope given by the convex hull of the external data for $k=1$).  Projectively $(YAB)$ is a plane and the intersection with this polytope is just a pentagon on this plane. And the point $Y$ on $(YAB)$ is forced to lie inside this pentagon! Note this also suggests a different way of expressing the loop integrand/amplituhedron form than the usual one coming from BCFW triangulation. Traditionally for this case we would write the form as ``4-form $\times$ 4-form": one 4-form (for the $Y$ dependence corresponding to the ``R-invariant"), multiplied by another 4-form for the loop $(AB)$. In the new picture, it is more naturally expressed as ``6 form $\times$ 2 form", where the 6-form is the canonical form for $(YAB)$ in the $m=2$ amplituhedron, and the ``2-form" is the one for $Y$ on $(YAB)$ inside the aforementioned pentagon. 

As another example, let us look at the case of $k=0,n=5,L=2$. In the old definition, we look at two $(2 \times 5)$ ``$D$" matrices $D_{1,2}$. We first have to demand that both $D_{1,2}$ are positive (which means that $AB_1$ and $AB_2$ are in the usual 1-loop (same as $m=2$ tree) amplituhedron), together with the requirement that all the ordered $(4 \times 4)$ minors of the $(4 \times 5)$ matrix stacking $(D_1,D_2)$, are positive. This certaily implies that $\langle (AB)_1 (AB)_2 \rangle>0$, but seems to demand even more. However, our new claim is that, once $(AB)_1, (AB)_2$ are in the 1-loop amplituhedron, then demanding $(AB)_1 (AB)_2 > 0$ is enough to enforce being in the 2-loop amplituhedron. 

It is straightforward to check this picture by computing the full 2-loop amplitude, but in order to illustate the methods in a simpler non-trivial example, let us compute a ``cut" of the 2-loop $n=5$ amplitude where $\langle A B  1 2 \rangle \to 0$, and $\langle CD 3 4 \rangle = 0, \langle C D 4 5 \rangle = 0$. For simplicity we use positive data where $Z_5 = -Z_4 + Z_3 - Z_2 + Z_1$ and also normalize  $\langle Z_1 Z_2 Z_3 Z_4 \rangle = 1$. Given the 3-term triangulation of the 1-loop amplituhedron, it is easy to see that on this cut $CD$ only belongs to a single cell, and can be put in the form $C = Z_3 + u Z_4, D= Z_4 + v Z_5$ with $u,v>0$. 
There are two one-loop cells which cover $\langle A B 1 2 \rangle =0$; so just demanding that $(CD),(AB)$ are in the 1-loop amplituhedra tells us we can parametrize 
\begin{equation}
C = Z_3 + u Z_4, D = Z_4 + v Z_5, A = Z_1 + x Z_2, B=\left\{\begin{array}{c} Z_3 + y Z_2 + z Z_4 \\ -Z_1 + \alpha Z_4 + \beta Z_5 \end{array} \right\}; u,v,x,y,z,\alpha,\beta>0
\end{equation}
Now in each of these cells, we have the additional condition that $\langle A B C D \rangle > 0$. For instance in the first cell, we have 
\begin{equation}
\langle A B C D \rangle = (1+v) y + u v(1 + x + y) - v(1 + x) z > 0
\end{equation}
and thus we have the inequalities 
\begin{equation}
x,y,u,v>0, 0<z<\frac{(1+v) y + u v (1 + x + y)}{v(1+x)}
\end{equation}
and the corresponding form is 
\begin{equation}
\Omega^{(1)} = \frac{dx}{x} \frac{dy}{y} \frac{du}{u} \frac{dv}{v} dz \left(\frac{1}{z} - \frac{1}{z - \frac{(1+v) y + u v (1 + x + y)}{v(1+x)}}\right)
\end{equation}
Exactly the same exercise for the second cell gives us the inequalities
\begin{equation}
x,\beta,u,v>0, 0<\alpha<\frac{x(1 + v +  v) + \beta(1+x)}{v(1+x)}
\end{equation}
and the form 
\begin{equation}
\Omega^{(2)} = \frac{dx}{x} \frac{d \beta}{\beta} \frac{du}{u} \frac{dv}{v} d \alpha \left(\frac{1}{\alpha} - \frac{1}{\alpha - \frac{x(1 + v +  v) + \beta (1+x)}{v(1+x)}}\right)
\end{equation}
Now we simply add the two forms. Of course since we have used different variables to parametrize $B$ in the two cells, we have to make the co-ordinate change between them. We can always expand $B$ as either $B=Z_3 + y Z_2 + z Z_4$ or as $B = -Z_1 + \alpha Z_4 + \beta Z_5$ (of course in general with no sign restriction on $y,z,\alpha,\beta$). Mathcing $(AB)$ in these two co-ordinates gives us the relationship between the parameters as $\alpha = - x/(1 + x + y), \beta = x(1 + z)/(1 + x + y)$.
Inserting this into the expression for $\Omega^{(2)}$ and adding $\Omega^{(1)}$ gives an expression for $\Omega = \Omega^{(1)} + \Omega^{(2)}$:
\begin{equation}
\Omega = dx dy dz du dv \frac{u v(1+ x + y)(1 + x + y + z + x z) + y(1 + x + y +v(1 + x + y) + z(1+x)}{u v x y (1 + x + y) z (1+z) ((1+v) y + u v (1 + x + y) - v(1+x) z)}
\end{equation}
This expression precisely (and highly non-trivially) matches the corresponding cut of the 2-loop amplitude. 

Our new description of the full amplituhedron for both trees and loops now has a satisfyingly strong resonance with three central aspects of scattering amplitude physics. The ``$m=0$" part of the geometry is about understanding the geometry of mutual positivity between loops $\langle (AB)_\gamma (AB)_\rho\ra > 0$; this is present even for the simplest case of $k=0,n=4$, and is associated with the physics of the universal IR divergences and the cusp-anomalous dimension. The ``$m=4$" part of course has to do with the physics of tree amplitudes. Finally the ``$m=2$'' part is the physics of the leading quantum corrections. 

\section{The Amplituhedron in Twistor Space}

As we have remarked, it is striking that our new picture of the amplituhedron makes reference only to what the configuration of $Z$'s looks like after projecting the $(k+m)$-dimensional ${\cal Z}$ data through $Y$. For the case of relevance to scattering amplitudes with $m=4$, this means that everything can be described as a property of the configuration of bosonic momentum-twistor data! This is pleasing, since from a physical point of view, while the $m=4$-dimensional momentum-twistors have a manifest importance as specifying the external kinematical data, the introduction of the extra $k$ components of the ${\cal Z}$'s, and the $k$-plane $Y$, is more mysterious, related to a ``bosonization" of the supersymmetry. This structure is needed since the canonical amplituhedron form lives in $Y$ space, and the super-amplitude is extracted from it \cite{Arkani-Hamed:2013jha}. But given that our new definition of the amplitude seems to make reference only to the $m$-dimensional space, it would be very pleasing if the geometry of the amplituhedron as well as the superamplitude could be directly associated with the $m$-dimensional space, without ever referring to $Y$ or the underlying $(m+k)$ dimensional ${\cal Z}$ data. 

This is very easy to do. We are working with the configuration space ${\cal M}(m,n)$ of $n$ vectors $Z_a$ in $m$ dimensions. 
Let us define the subspace of ${\cal M}(m,n)$ where the configuration has the correct ``winding" or ``flip" pattern we have discussed earlier appropriate to some $k$ as ${\cal W}(k,m,n) \subset {\cal M}(m,n)$. Now, the space ${\cal M}(m,n)$ is $m \times n$ dimensional, and the subset ${\cal W}(k,m,n)$ is clearly a top-dimensional subspace and is also $m \times n$ dimensional. On the other hand, the amplituhedron is $k \times m$ dimensional and has lower dimension. We would  thus like to identify subspaces in ${\cal M}(m,n)$ that can be obtained from some fixed $(m+k)$ dimensional data ${\cal Z}_a$ by projecting though some $k$-plane $Y$. 

But this is both natural and trivial. Suppose we begin with some fixed set of vectors $Z_{* a}$ that give us a point in ${\cal M}(m,n)$. We can think of this as giving as a fixed $m$-plane $Z_*$ in $n$ dimensions. Now, let us consider the affine subspace which are linear translates of this $m$-plane, by translating in directions lying in some fixed $k$-plane $\Delta$ in $n$ dimensions. 

$$
\includegraphics[scale=.55]{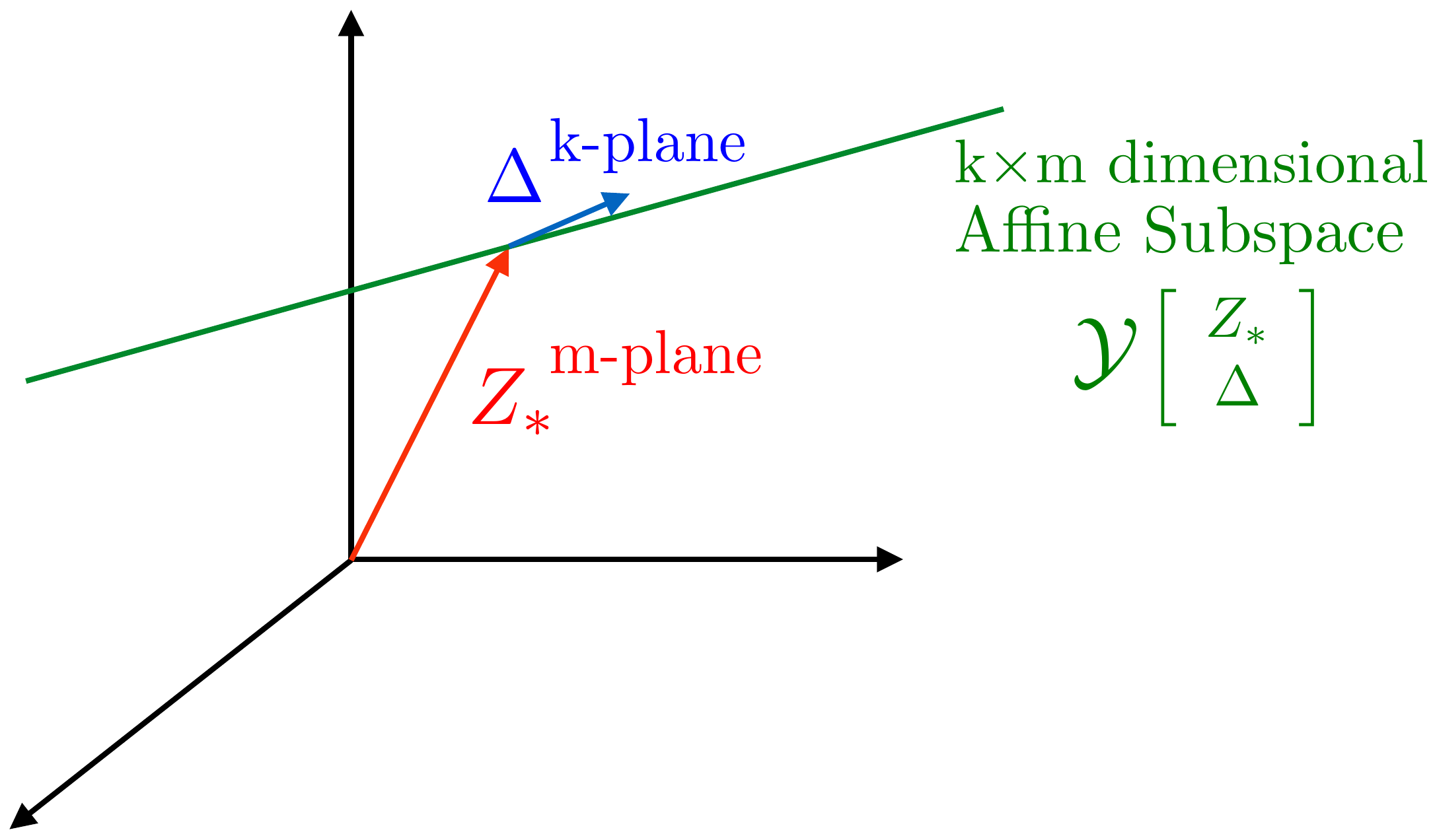}
$$
In equations, we look at the space of all $Z_a^i$ that can be obtained starting from $Z_{* a}^i$ and translating in the direction of $\Delta^\alpha_{a}$, i.e. all $Z_a^i$ of the form 
\begin{equation}
Z_a^i = Z_{* a}^i + y^{i}_{\alpha} \Delta^{\alpha}_{a}
\end{equation}
(Here $i=1, \cdots, m$ is the vector index on the $m-$dimensional space). 

Note that such a subspace is specified by giving a $(k+m) \times n$ matrix of data, 
\begin{equation}
{\cal Z}^I_a = \left(\begin{array}{c} Z^i _{* a} \\ \Delta^\alpha_a \end{array} \right)
\end{equation}
which is what we think of as ``fixed external data" in the usual amplituhedron story. Here the index $I$ runs from $I=1, \cdots, (k+m)$; we can think of the first $m$ components as corresponding to the $i$ indices and the last $k$ components as corresponding to the $\alpha$ indices. Furthermore, the $Z_a$ above are precisely what we get by projecting the $(k+m)$-dimensional ${\cal Z}_a$ data through the $k$-plane $Y_\alpha^I$ in $(k+m)$ dimensions, where
\begin{equation} 
Y_{\alpha}^{I=i} = - y_\alpha^i, \, Y_{\alpha}^{I=\beta} = \delta_\alpha^\beta
\end{equation}

Thus from the $m$-dimensional point of view, specifying $(k+m)$-dimensional data is actually picking out a particular translation of the subspace of ${\cal M}(m,n)$ from a $k\times m$ space of possible translations.  We can refer to this
translation of $M(m,n)$ as the affine subspace ${\cal Y}[{\cal Z}]$. 

For any of these affine subspaces, we can look at the part of the subspace which is compatible with the correct winding/flip pattern, and gives us the $m$-dimensional characterization of the amplituhedron ${\cal A}_{m,k,n}[{\cal Z}]$: 
\begin{equation}
{\cal A}_{m,k,n}[{\cal Z}] = {\cal Y}[{\cal Z}] \cap {\cal W}[m,k,n]
\end{equation}

In this picture, there is one last vestige of the $(k+m)$ dimensional picture--we must demand that this affine subspace be ``positive" in the sense that all the ordered minors of the ${\cal Z}$ matrix are positive. It is interesting to ask the extent to which we can remove even this restriction. To begin with we can ask the following obvious question. Suppose we have some $m$-dimensional configuration of $Z$'s satisfying the right winding condition. Is it guaranteed that we can think of having obtained this data by projecting ${\it positive}$ $(k+m)$-dimensional data ${\cal Z}$ through some $k$-plane $Y$? Said more prosaically: given some $(m \times n)$ matrix of $Z_a$'s that satisfies the winding/flip criteria, can we always add $k$ more rows so that the resulting $(k+m) \times n$ matrix is positive? 

While we do not have a general proof of this statement, we suspect that the answer is likely ``yes". A sketch of an approach to a proof might be the following, setting $m=2$ for simplicity.  We'd like to show that whatever $m=2$ dimensional data we have with correct winding, we can uplift it to positive $(k+2)$ dimensional data. Now,  if we have two different collections of $Z_i$ with the same orientation for the windings for both curves, then we should be able to smoothly deform one configuration into the other. If the orientations of each segment $(ii+1)$ are also the same, then it seems plausible that such a deformation can be generated by a combination of elementary moves on the vertices $Z_i$: rescaling $Z_i$  by a positive constant, or moving $Z_i$ in the direction of either of its neighbors; i.e. a series of operations of the  form $Z_i \to x_i Z_i + x_{i-1} Z_{i-1} + x_{i+1} Z_{i+1}$. But these moves on the projected $Z_i$ follows under projection from exactly the same operation on the ${\cal Z}$'s in the $(k+2)$-dimensional space ${\cal Z}_i \to x_iZ_i + x_{i-1} Z_{i-1} + x_{i+1} Z_{i+1}$, and this operation preserves the positivity of the ${\cal Z}$ data. 

Finally this picture clearly extends to all loop orders. For the case $m=4$ of relevance to scattering amplitudes, aside from the $Z_a$ we also have $L$ planes $(AB)_\alpha$ which are 2-planes in 4-dimensions. We can define ${\cal W}(m=4,k,n;L) \subset {\cal M}(m=4,k,n) \times (AB)^L$ as that subset that has the correct winding properties at loop level. Then, the loop-level amplituhedron is 
\begin{equation}
{\cal A}_{m=4,k,n}[{\cal Z}] = \left({\cal Y}[{\cal Z}] \times (AB)^L \right) \cap {\cal W}[m=4,k,n;L]
\end{equation}

\section{(Super)-Amplitudes As Differential Forms on Twistor Space}

Having seen the $m$-dimensional image of the amplituhedron without any reference to $Y$, let us go further and discuss how to think about the canonical form and the scattering (super)-amplitude in an intrinsically $m$-dimensional way---as we will see the super-amplitude is {\it literally} a degree $m \times k$ differential form on the configuration space ${\cal M}(m,n)$ of the $m$-dimensional $Z_a$'s. Before showing how this works in generality, let's start with a simple example familiar from the simplest scattering amplitudes with $m=4,k=1$, which are built out of the well-known ``$R$-invariants". Let's first describe the $R$-invariant in standard terms, as a super-amplitude, which we can write as 
\begin{equation}
(1 2 3 4 5) = \frac{\delta^4\left(\langle 1 2 3 4 \rangle \eta_5 + {\rm cyclic} \right)}{\langle 1 2 3 4 \rangle \cdots \langle 5 1 2 3 \rangle}
\end{equation}
Now in the language of the amplituhedron, we instead talk about a 4-form with logarithmic singularities on $Y$ space, that is 
\begin{equation}
\Omega(Y, {\cal Z}_a)  =  d_Y {\rm log}\frac{[\, Y 1 2 3 4 ]\,}{ [\,Y 5 1 2 3 ]\,} \cdots d_Y {\rm log} \frac{[\, Y 4 5 1 2 ]\,}{[\, Y 5 1 2 3 ]\,}  =  \frac{[\,Y d^4 Y ]\, [\, 1 2 3 4 5 ]\,^4}{[\, Y 1 2 3 4 ]\, \cdots [\, Y 5 1 2 3 ]\,}
\end{equation}
Here the subscript on $d_Y$ is to remind us that we are to take the external data as fixed, with the differentials acting on $Y$. Starting from this form there is a simple prescription for extracting the superamplitude, but we will present a more direct and striking connection. Note that of course all the brackets occurring as arguments of the dlog's above contain $Y$; thus we can interpret them all as 4-brackets on the space of momentum-twistors obtained when projecting through $Y$. It is then very natural to look at a 4-form, not on $Y$ space, but on momentum-twistor space, as 
\begin{equation}
\Omega(Z_a) =  d_Z{\rm log}\frac{\langle  1 2 3 4 \rangle}{\la 5 1 2 3 \rangle} \cdots d_Z {\rm log} \frac{\langle  4 5 1 2 \rangle}{\langle  5 1 2 3 \rangle} 
\end{equation}
Now, one can directly verify that this form can be re-written as 
\begin{equation}
\Omega(Z_a) =  \frac{\delta^4\left(\langle 1 2 3 4 \rangle dZ_5 + {\rm cyclic} \right)}{\langle 1 2 3 4 \rangle \cdots \langle 5 1 2 3 \rangle}
\end{equation}
Note that remarkably, this is precisely the $R$-invariant, if the (anti-commuting) super-variables $\eta_a^I$ are replaced by the differentials $\eta_a^I \to dZ_a^I$! We will shortly understand why this happens on general grounds, but let us first make some general comments. 

Suppose we have some $m \times k$ form on the Grassmannian, and let us consider the pull-back of this form to some $m \times k$ dimensional subspace of $G(k+m,n)$. We can describe this by some $C_{\alpha a}(x_1, \cdots, x_{m \times k})$. Now, consider those $k$-planes that are constrained by being orthogonal to some $m$-plane, $Z$; this will generically intersect the subspace in points; concretely we are just saying that the equations 
\begin{equation}
C_{\alpha a}(x_i) Z_a^I = 0,  \textrm{ has solutions } x_i = x_i(Z_a)
\end{equation}
We would like to push forward any form from the Grassmannian onto $Z_a$ space, in other words we would like to re-write the measure 
$dx_1 \cdots dx_{m \times k}$ in terms of the wedge products of $m \times k$ of the $dZ_a$'s. The result is simple; we will show that 
\begin{eqnarray}
dx_1 \cdots dx_{m \times k} &=& \int d y_1 \cdots d y_{m \times k} \, \delta^{m \times k}\left[C_{\alpha a}(y) Z_a \right] 
\delta^{m \times k} \left[C_{\alpha a} d Z_a \right] \nonumber \\ & = &  \int d y_1 \cdots d y_{m \times k} \, \delta^{m \times k|m \times k}\left[C_{\alpha a}(y) {\bf Z}_a \right] \, {\rm with \, } \eta_a^I \to dZ_a^I
\end{eqnarray}
The proof is easy. Let's start with taking the differential of $C_{\alpha a} Z_a^I = 0$, to find 
\begin{equation}
\frac{\partial C_{\alpha a}}{\partial x_i} Z_a^I dx_i = - C_{\alpha a}(x) dZ_a^I
\end{equation}
Taking the $m \times k$ power of both sides we find 
\begin{equation}
{\rm det}_{\{\alpha I, i\}} \left(\frac{\partial C_{\alpha a}(x)}{\partial x_i} Z_{a}^I \right) dx_1 \cdots dx_{m \times k} = \delta^{m \times k}\left(C_{\alpha a} dZ_a\right)
\end{equation}
so that 
\begin{eqnarray}
dx_1 \cdots dx_{m \times k} &=& \left[{\rm det}_{\{\alpha I, i\}} \left(\frac{\partial C_{\alpha a}(x)}{\partial x_i} Z_{a}^I \right)\right]^{-1} \times
\delta^{m \times k}\left(C_{\alpha a}(x) d Z_a\right) \nonumber \\ &=& \int d y_1 \cdots d y_{m \times k} \, \delta^{m \times k}\left[C_{\alpha a}(y) Z_a \right] 
\delta^{m \times k} \left[C_{\alpha a} d Z_a \right]
\end{eqnarray}
as desired. 

Thinking of the canonical amplituhedron forms instead as $m \times k$ forms on the $m$-dimensional space of $Z_a$ data exposes some remarkable relationships between forms that are not evident from the conventional $Y$-space picture. 
Let us return to the $m=2$ amplituhedron for which we gave a triangulation and determined the form in section 7. We can re-interpret these as forms on the space of $2$-dimensional vectors $Z_a$. For $k=1$, from the triangulation $\sum_i (1 i i+1)$ of the polygon we have
\begin{equation}
\Omega_{k=1,m=2} = \sum_i \frac{(\langle Z_1 Z_i \rangle dZ_{i+1} - \langle Z_1 Z_{i+1} \rangle dZ_i + \langle Z_i Z_{i+1}\rangle dZ_1)^2}{\langle Z_1 Z_{i} \rangle \langle Z_1 Z_{i+1} \rangle \langle Z_i Z_{i+1} \rangle}
\end{equation}
But there is now a beautifully simple expression for the $2 \times k$ form for any $k$, we have 
\begin{equation}
\Omega_{k,m=2} = \frac{\Omega_{k=1,m=2}^k}{k!}
\end{equation}

This understanding of the scattering amplitude as a differential form obviously extends to loop level as well. In addition to the external twistor data $Z_a$, we also have $L$ 2-planes $(AB)_\alpha$, and 
we have a $4 \times (k+L)$ form on $\{Z_a,(AB)_\alpha\}$ space. Note that in this setting ``the loop integrand" is just one component of the $4(k+L)$ form. For instance, even the simplest $n=4$ 1-loop amplitude corresponds to the 4-form 
\begin{eqnarray}
d {\rm log} \frac{\langle A B 1 2 \rangle}{\langle A B 1 3 \rangle} \cdots d {\rm log} \frac{\langle A B 1 4 \rangle}{\langle A B 1 3 \rangle} & = & \\ 
\frac{\langle A B d^2 A \rangle \langle A B d^2 B \rangle \langle 1 2 3 4 \rangle^2}{\langle A B 1 2 \rangle \cdots \langle A B 1 4 \rangle} & + \cdots + & \frac{\langle A B 1 2 \rangle^2 \langle A B dZ_3 dZ_3 \rangle \langle A B dZ_4 dZ_4 \rangle + \cdots}{\langle A B 1 2 \rangle \cdots \langle A B 1 4 \rangle} 
\end{eqnarray}
The first terms, where all 4 $d$'s hit $(AB)$, is the familiar 1-loop integrand; then we have terms with a mixture of $d$'s hitting the $(AB)$ and the $Z_a$, and the last term, where the $d$'s hit only the $Z_a$. 

Our new picture of scattering amplitudes as differential forms is very satisfying. The ``super"-part of superamplitudes has always presented an obstruction between linking properties of the integrand on the one hand, and the final integrated amplitudes on the other. 
In particular, recent years has seen a fascinating emergence of cluster algebra structure in the polylogarithms found in ${\cal N}=4$ SYM amplitudes---the arguments of the polylogs are expressed as cross-ratios of momentum twistor data \cite{Hodges:2009hk} naturally associated with cluster algebras \cite{Goncharov:2010jf,Golden:2013xva} for the external kinematical data in ${\cal M}(4,n)$. This has long cried out for a link with the positive Grassmannian/amplituhedron structure at the level of the integrand, but the ``$\eta"$'s in superamplitudes obscure this connection. The bosonization of the integrand afforded by the amplituhedron improves the situation, but leaves us with external data that is $(4+k)$ dimensional while obviously the cluster structure in integrated results only knows about 4-dimensional momentum twistor data. 
But finally with the new picture of amplitudes as forms, integrand and amplitudes are on a fully equal footing, depending on the same variables. As we have seen, however, the ``positive geometry" associated with external data in the 4-dimensional space is not merely ``positivity", but involves further  combinatorial/topological ``winding/flip" criteria. It will be fascinating to understand how these may be reflected in the transcendental functions appearing after loop integration.

\section{Parity}

Parity is a fundamental symmetry of scattering amplitudes which is conventionally completely obscured in momentum-twistor space. The bosonic action of the symmetry is easy to see: given momentum twistors $Z_a^I$, we have the parity conjugates $W_{a I}$ which are the planes $(Z_{a-1} Z_a Z_{a+1})$; with an additional factor of $(-1)$ for $a=1$ and $a=n$. Now, for the full scattering amplitudes labeled by $(n,\widehat{k})$, parity interchanges $\widehat{k} \leftrightarrow \widehat{(n-k)}$; but in terms of $k$ this is the rather more peculiar looking interchange of $k \leftrightarrow (n-k-4)$, which presumably reflects a symmetry $k \leftrightarrow (n - k - m)$ for general $m$. As we will now see, our ``winding" picture gives a beautifully simple understanding of these symmetries.  

The are in fact two different $\mathbb Z_2$ symmetries that in concert give us the physical parity. The first one is extremely simple but already shows strikingly why a $k \leftrightarrow (n - m - k)$ symmetry should be expected. Suppose we simply change $Z_a \to (-1)^{a} Z_a$ i.e. we flip the sign of ever other $Z$. Now consider (e.g. for $m=2$) the sequence $\{\langle 1 2\rangle, \langle 1 3 \rangle, \cdots \langle 1 n  \rangle\}$.
Note that the number of possible positions of sign-flips of this sequence is $(n-2)$, and is $(n-m)$ for general $m$. Now, obviously if two consecutive signs agree before this transformation, they will disagree afterwards, and vice-versa. So this changes the number of sign flips from $k \to (n- m - k)$!

This is a rather trivial $\mathbb Z_2$ which knows nothing about the $W_a$. There is a more non-trivial fact featuring the $W_a$: if the $Z$'s are in the amplituhedron, i.e. that $\langle ii+1 jj+1 \rangle > 0$ and the sequence $\{\langle 1 2 3 4 \rangle, \cdots, \langle 1 2 3 n \rangle\}$ has $k$ sign flips, then the $W$'s are {\it also} in the amplituhedron with the same value of $k$!

First note that as long as $\langle Z_i Z_{i+1} Z_{j} Z_{j+1}\rangle > 0$, then also $\langle W_i W_{i+1} W_j W_{j+1}\rangle>0$, since 
\begin{equation}
\langle W_i W_{i+1} W_j W_{j+1} \rangle = \langle Z_i Z_{i+1} Z_j
Z_{j+1} \rangle \langle Z_{i-1} Z_i Z_{i+1}Z_{i+2}\rangle \langle Z_{j-1} Z_j Z_{j+1} Z_{j+2} \rangle
\end{equation}
Note that for this conclusion we don't have to assume that all the minors of these $Z$'s are positive (which they aren't!), only that the minors of the form $\langle aa+1 bb+1 \rangle>0$. 

Now, we only have to show that the sequence $\{\langle W_1 W_2 W_3 W_4 \rangle, \langle W_1 W_2 W_3 W_5 \rangle, \cdots \langle W_1 W_2 W_3 W_n \rangle\}$ has $k$ sign flips. A short computation of these four brackets turns this into the pretty statement that the sequence 
\begin{equation}
\{ \langle 1 (234) \rangle,  \langle 1 (345) \rangle, \cdots, \langle 1 (n-2 \, n-1 \, n) \rangle \}  \textrm{ has $k$ sign flips}
\end{equation}
 This statement is easy to prove. Let us consider the following sequences of minors 
\begin{equation}
\begin{array}{ccccc} \langle 1 2 3 4 \rangle & \langle 1 2 3 5 \rangle & \langle 1 2 3 6 \rangle & \cdots&  \langle 1 2 3 n \rangle \\  \langle 1 2 3 4 \rangle & \langle 1 3 4 5 \rangle & \langle 1 3 4 6 \rangle & \cdots & \langle 1 3 4 n \rangle \\ \langle 1 2 3 4 \rangle & \langle 1 3 4 5 \rangle & \langle 1 4 5 6 \rangle &\cdots  & \langle 1 4 5 n \rangle \\ 
\vdots & \vdots & \vdots & \cdots & \vdots \\ \langle 1 2 3 4 \rangle & \langle 1 3 4 5 \rangle & \langle 1 4 5 6 \rangle & \cdots & \langle 1 (n-2 \, n-1 \, n) \rangle \end{array} 
\end{equation}
The first row is our usual sequence $\{\langle 1 2 3 i \rangle \}$, which has $k$ sign flips. The second row has the same first entry as the first row, and thereafter is of the form $\langle 1 3 4 i \rangle$. The third row has the same first two entries as the second row, and is thereafter of the form $\langle 1 4 5 i \rangle$, and so on. Now it is easy to see that the number of sign flips of the $i$'th and $(i+1)$'st rows must be the same. The first parts of the two rows coincide; thereafter the argument is exactly the same as what we used to show that the number of sign flips for the $m=2$ amplituhedron is independent of the point we project through, namely, that by Plucker, we know that either there are no sign flips in successive slots, or both flip, or if the top row flips, the next slot where a flip  occurs just in one row, it must occur in the bottom row. Since we know that all the last entries are of the form $\langle ii+1 1 n \rangle$ and thus have a fixed sign, this means that the number of sign flips must be equal. In this way we work our way from the top to bottom rows, and conclude that the sequence $\{\langle 1(234) \rangle, \langle 1 (345) \rangle, \cdots, \langle 1 (n-2 \, n-1 \, n) \rangle$ has $k$ sign flips, as desired. 

The statement of parity at loop level is more interesting. Let's work at one-loop to begin with. We know that when we project through $AB$, the sequence $\langle A B 1 2 \rangle, \langle A B 1 3 \rangle, \cdots, \langle A B 1 n \rangle$ should have $k+2$ sign flips. Now, we would like to see what happens when we dualize  the $Z_i$ to $W_i$; our claim is that loop-level parity is the statement that the sequence 
\begin{equation}
\{ \langle A B W_1 W_2 \rangle, \langle A B W_1 W_3 \rangle \cdots, \langle A B W_1 W_n \rangle \} \textrm{ has }  k  \textrm{ sign flips} 
\end{equation}
In general, we can expand
\begin{equation}
\langle AB W_1 W_j \rangle = \langle AB n 1\rangle  \langle 2 j-1 j j+1 \rangle - \langle AB n 2 \rangle  \langle 1 j-1j j+1 \rangle +
\langle AB12 \rangle \langle n j-1 j j+1 \rangle.
\end{equation}
If we want to write this back with $Y$'s, we have to add a $Y$ to both sets of brackets, $[\, YAB n 1 ]\,  [\, Y2 j-1 j j+1 ]\, - [\,YAB n 2 ]\, [\, Y1 j-1j j+1 ]\, + [\, YAB12 ]\, [\, Yn j-1 j j+1 ]\,$. The claim is that this sequence should have $k$ sign flips. For $k=0$, this is the statement that all $\langle A B W_1 W_i \rangle$ are positive, a statement we will say more about in section 14. We don't have a proof for general $k$, though we have checked these statements numerically for a large range of $k,n,L$.

\section{Different winding sectors, ${\cal M} \times \overline{{\cal M}}$ and Correlation Functions}

Given that the amplituhedron has maximal winding, it is natural to ask whether there is any meaning to sectors with different winding/flip patterns.  Let us start again with the case of $m=2$, $k=2$. The amplituhedron corresponds to winding number 1, but is there some meaning to the sector where we still have $[Y i i+1]>0$ but where we have winding number 0? The interpretation of the $m=2$ amplitudes as the 1-loop integrand for the MHV amplitudes suggests an obvious candidate. We know that by parity we can replace $Z_a$ with $W_a = (Z_{a-1} Z_a Z_{a+1})$; doing this takes us from the integrand for MHV amplitudes to that for $\overline{{\rm MHV}}$ amplitudes. So it is natural to conjecture that the canonical form associated with winding $0$ sector corresponds to the 
$\overline{{\rm MHV}}$ 1-loop integrand. We have verified empirically that this is correct, by identifying the $\overline{{\rm MHV}}$ integrand with the canonical form with logarithmic singularities on the minimally winding space. The forms are of course different, for instance for $n=5$ we have 
\begin{equation} 
{\cal M}_{{\rm MHV}} = \frac{\la AB(512)\cap(234)\ra\la 3451\ra -\la AB51\ra\la 1234\ra\la 2345\ra  - \la AB34\ra\la 4512\ra\la 5123\ra}{\la AB12\ra\la AB23\ra\la AB34\ra\la AB45\ra\la AB51\ra}
\end{equation}
while 
\begin{equation}
{\cal M}_{{\rm \overline{MHV}}} = \frac{\la AB13\ra\la 2345\ra\la 4512\ra - \la AB51\ra\la 1234\ra\la 2345\ra - \la AB34\ra\la 4512\ra\la 5123\ra}{\la AB12\ra\la AB23\ra\la AB34\ra\la AB45\ra\la AB51\ra}
\end{equation}

It is interesting to note a feature of the geometry also reflected in the forms. The winding number 1 and 0 regions are {\it almost} disjoint, in the sense that they don't touch on co-dimension one boundaries. Only when we go to higher-dimensional boundaries that correspond to collinear regions do the two regions touch. This is reflected in the forms: while both forms have the same ``physical poles", the forms are different, and the residues on the co-dimension one boundaries are also different. But upon taking enough residues and going to high enough co-dimension boundaries, the forms match when the shared boundaries match. 

We can continue in this way to discuss any number of loops, still with $k=0$. When projecting through each $(AB)_i$, we either get winding number 0 or 1. If we define the all-loop integrand for MHV and $\overline{{\rm MHV}}$ amplitudes  to be ${\cal M}$, written in a loop expansion as
\begin{equation}
{\cal M} = 1 + g^2 {\cal M}_1 + g^4 {\cal M}_2 + \cdots, \, \overline{{\cal M}} = 1 + g^2 \overline{{\cal M}}_1 +  g^4 \overline{{\cal M}}_2 + \cdots
\end{equation}
then 
\begin{equation} 
{\cal M} \overline{{\cal M}} = 1 + g^2 ({\cal M}_1  +  \overline{{\cal M}}_1) + g^4 ({\cal M}_2 + \overline{{\cal M}}_2 +  {\cal M}_1 \overline{{\cal M}}_1 + \overline{{\cal M}}_1 {\cal M}_1) + \cdots
\end{equation}
At each loop order, we are adding over all the possible winding numbers, and thus ${\cal M} \times \overline{{\cal M}}$ is naturally decomposing the space defined simply by the boundary inequalities 
\begin{equation}
\langle (A B)_j i i+1 \rangle > 0, \langle (AB)_i (AB)_j \rangle > 0 
\end{equation}
into the pieces with different winding numbers. It is interesting to note that at 1-loop there is a well-defined form with logarithmic singuarties on these boundaries. Interestingly, it does not correspond to the parity even {\it sum} of the MHV and $\overline{{\rm MHV}}$ integrands,but to the parity odd difference between them, which vanishes for $n=4$ but is non-vanishing for higher $n$.  

It is interesting that this space defined by the ``obvious physical boundaries" inequalities has a nice physical interpretation. Indeed we began our investigations in this paper by noting that the space we get simply from imposing the obvious physics boundaries 
\begin{equation}
[Y i i+1 j j+1] > 0, \, [Y (A B)_\alpha i i+1] > 0, [Y (AB)_\alpha (AB)_\beta]>0
\end{equation}
does not give us the amplituhedron, but it may find a natural meaning related to the ``square" of the amplitude, and simplified even further, to correlation functions. Indeed recent years have seen a beautiful connection between amplitudes and the light-like limit of stress-tensor correlators in ${\cal N}=4$ SYM. For MHV amplitudes, it is natural to think of the lines $(Z_i Z_{i+1})$ and the loops $(AB)$ on the same footing; we can indeed think of a collection of lines ${\cal L}_{1, \cdots, n+L}$. Going to the light-like limit simply picks $n$ of these lines ${\cal L}_i$ out and asks ${\cal L}_a$ to intersect ${\cal L}_{a+1}$ cyclically. The correlation function itself is however a fully permutation invariant function of all the ${\cal L}_i$. It is therefore tempting to associate the geometry $\langle {\cal L}_i {\cal L}_j\rangle > 0$ with the correlation function. For general $k$, we can do the same thing; we consider some number of $(k+2)$-planes in $(k+4)$ dimensions $(Y {\cal L}_i)$ for $i=1, \cdots, (n+L)$, which overlap on $Y$, in such a way the planes are ``mutually positive"
\begin{equation} 
[Y {\cal L}_i {\cal L}_j]>0
\end{equation}
This is a perfectly well-defined space, but the crucial question is, how can we associate a form with this geometry to reproduce the correlation functions? An inspection of the correlators themselves shows that they do not have logarithmic singularities---upon taking residues we encounter double-poles that ruin the logarithmic property. It would be fascinating to nonetheless find {\it some} way of associating a form with this space. One obvious strategy is simply to ask the form to become logarithmic in the lightlike limit, where we know that the geometry does decompose into different winding sectors with well-defined forms associated with the square of the amplitude $M^2$. But it would be much more satisfying if this could be done more intrinsically; see \cite{Eden:2017fow} for some interesting attempts along these lines. 
Obviously any such picture must contain all the intricate information associated with the topology of the amplituhedron. 

There are a number of other interesting objects closely related to the amplitudes and the amplituhedron canonical form. For instance we saw that for $m=2$, the canonical form for any $k$ is the $k$'th power of the form for $k=1$; while this doesn't hold true for $m=4$, we have nonetheless observed this ``$k$'th power form" is interesting, for instance it non-trivially has only simple poles. Even more interestingly, at loop level we have the natural ``ratio function" which is the ratio ${\cal M}_{n,k,L}/{\cal M}_{n,k=0,L}$. Might any of these objects be associated with different winding sectors?

\section{A ``Dual" of the Amplituhedron}

Continuing in the vein of exploring the significance of different winding/flip sectors, it is natural to ask about a natural counterpart to the amplituhedron: what space do we define if, in projecting through some $k^\prime$-plane, the resulting data has ``minimal" winding?  It  is in particular natural to ask this for projecting through $k^\prime = m$-dimensional planes $\tilde{Y}$ in $(k+m)$ dimensions; the dimensionality of $\widetilde{Y}$ space is $m \times k$, the same as the amplituhedron.  Now, ``minimal flips" has a meaning; as we saw in our discussion of the $k=0$ amplituhedron, the positive Grassmannian itself gives us configurations of ``zero flips" via projection down to one dimension. So, it is natural to define the subspace of $m$-planes in $(k+m)$ dimensions which satisfy 
\begin{equation}
\langle \widetilde{Y} {\cal Z}_{a_1} \cdots {\cal Z}_{a_k} \rangle > 0 \,\,\, {\rm for} \,\, a_1<\cdots<a_k
\end{equation}
A related motivation for defining this space is simply the following. We may have extremely naively thought that starting with positive $(k+m)$-dimensional data and projecting through $Y$ in the amplituhedron would have left us with positive data. As we have seen this is wrong, but it is natural to ask what planes $\widetilde{Y}$ do have such a property. Note that in the language of section 5.1, the $\widetilde{Y}$ are ``positive projections" ${\cal P}_{m \to m^\prime = 0}$ down to $m^\prime = 0$. The space of $\widetilde{Y}$'s of this form is not empty, easy examples are afforded by looking at a matrix of positive ${\cal Z}_a$ data that has the form of the ``moment curve" 
\begin{equation} 
\left(\begin{array}{ccc} 1 & \cdots & 1 \\ x_1 & \cdots & x_n \\ x_1^2 & \cdots & x_n^2 \\ \vdots & \vdots & \vdots \\ x_1^{k+m-1} & \cdots & x_n^{k+m-1} \end{array} \right) \, {\rm with} \, x_1 < \cdots < x_n
\end{equation}
All the ordered minors of this matrix are given by Vandermonde determinants and are positive for $x_1< \cdots < x_n$. 
But if we take $\widetilde{Y}$ to be an $m-$plane given by the bottom $m$ rows of this matrix, then projecting through $\widetilde{Y}$ will give us $k$ dimensional data that simply corresponds to the top $k$ rows of the matrix which are still positive. 

Note that the space of $\tilde{Y}$'s defined in this way has the property that 
\begin{equation}
\langle \widetilde{Y} \cdot Y \rangle > 0 \textrm{ for  all $Y$ in the amplituhedron}
\end{equation}

For the special case of $k=1$ for any $m$, the $\tilde Y$'s are $m$-planes in $m+1$ dimensions, which are points in the dual ${\mathbf P}^{m}$. Then the inequalities 
\begin{equation}
\langle \widetilde{Y} \cdot Z_a \rangle > 0
\end{equation}
are the equations defining a polytope in the dual ${\mathbf P}^{m}$, whose facets are the $Z_a$. So for $k=1$, this space can be identified with the dual of the (cyclic) polytope coming from the external data. The obvious extension to general $k$ gives one natural working definition for a dual of the amplituhedron, as described in \cite{Arkani-Hamed:2014dca,Arkani-Hamed:2017tmz}. 

This definition can naturally be extended to loops when $m=4$. At one-loop, we have $\widetilde{Y}$ which is a four-plane in $(4+k)$ dimensions; but we also have a 2-plane $\tilde{y}$ inside $\widetilde{Y}$; again this space of $4$-planes $\widetilde{Y}$ with a 2-plane $\tilde{y}$ inside it has the same dimensionality as the 1-loop amplituhedron for $m=4$. Requiring minimal winding when projecting through $\tilde{Y}$ and $\tilde{y}$ requires
\begin{equation}
\langle \widetilde{Y} {\cal Z}_{a_1} \cdots {\cal Z}_{a_k} \rangle > 0, \, \langle \tilde{y} {\cal Z}_{a_1} \cdots {\cal Z}_{a_{k+2}} \rangle > 0 
\end{equation}
At $L$-loop order, we have $L$ 2-planes $\tilde{y}_i$; in addition to the above constraints, we must also have that $\langle \langle \tilde{y}_i \tilde{y}_j \rangle \rangle > 0$,
where $\langle \langle a b c d \rangle \rangle$ represents contraction with the antisymmetric tensor on the 4-dimensional space defined by the 4-plane $\widetilde{Y}$. Said in the $(4+k)$-dimensional terms, this says 
\begin{equation}
\epsilon^{I_1 \cdots I_4 J_1 \cdots J_k} (\tilde{y}_i)_ {I_1 I_2} (\tilde{y}_j)_{ I_3 I_4} = \rho \epsilon^{I_1 \cdots I_4 J_1 \cdots J_k} \widetilde{Y}_{I_1 I_2 I_3 I_4}, \, {\rm with} \, \rho>0
\end{equation}

The space we have described certainly gives us {\it a} natural geometric ``dual" of the amplituhedron. As described in \cite{Arkani-Hamed:2014dca,Arkani-Hamed:2017tmz}, there are further motivations to find a dual amplituhedron; by analogy with the well-understood case of $k=1$, we can hope for a direct and intrinsic definition of the canonical form with logarithmic singularities on the amplituhedron expressed as an integral over the dual geometry. As already described in \cite{Arkani-Hamed:2017tmz} for the simplest case of $G_+(2,4)$, a direct extension of the analogy with $k=1$ already involves novel features not seen for polytopes. It will be interesting to see if the definition of the dual amplituhedron we have given will nonetheless end up playing an important role in determining the canonical form for both tree and loops.

\section{Cutting Out The Amplituhedron With Inequalities}

We began by asking whether there was a way of defining the amplituhedron analagous to the inequalities that cut out a polytope but immediately saw the obvious boundary inequalities are not enough. We have seen that this conditions must be supplemented by topological ones to determine the amplituhedron. Here we describe an alternative description which describes the amplituhedron purely by cutting it out with inequalities; we content ourselves with a brief description of these inequalities here, leaving a more complete investigation of this subject to future work. 

As we have seen, the ``winding" picture becomes natural when projecting through $Y$--we are focusing on the information that is contained in all the direction {\it not} spanned by the $k$-plane $Y$. Amusingly, the picture of inequalities is defined precisely in the opposite way, by looking at an interesting configuration of points {\it inside} $Y$. Let us start with $m=2$. We would like to identify points in the $k$-plane $Y$, which lives in $(k+2)$ dimensions. Just by dimension counting, a 3-plane in $(k+2)$ dimensions will intersect $Y$ in a point. But what natural 3-planes can we consider? Given the cyclic structure inherent in the set-up, it is natural to consider the 3-planes $({\cal Z}_{a-1} {\cal Z}_a {\cal Z}_{a+1})$, multiplied by some appropriate factors of $(-1)^{k-1}$ for $a=1,n$. These 3-planes intersect $Y$ in points that we will call $V_a$. Then, we claim that $Y$ is in the amplituhedron if and only if $[\, Y i i+1 ]\, > 0$, and also that the configuration of $n$, $k$-dimensional vectors $V_a$ is in the positive Grassmannian $G_+(k,n)$! 

Checking that $Y = C \cdot {\cal Z}$ satisfies this condition is interesting. These inequalities are satisfied due to somewhat magical positivity properties of the following ``determinants of minors". For instance for $k=2$ the claim is that as long as the ${\cal Z}$ data is positive, 
\begin{equation}
{\rm det} \left| \begin{array}{cc} [{\cal Z}_a {\cal Z}_{i-1} {\cal Z}_i {\cal Z}_{i+1} ] & [ {\cal Z}_a {\cal Z}_{j-1} {\cal Z}_j {\cal Z}_{j+1} ] \cr [{\cal Z}_b {\cal Z}_{i-1} {\cal Z}_i {\cal Z}_{i+1} ] & [{\cal Z}_b {\cal Z}_{j-1} {\cal Z}_j {\cal Z}_{j+1}]\, \end{array} \right|>0
\end{equation}
for any $a<b$ and $i<j$. This inequality follows non-trivially as a consequence of the positivity of the ${\cal Z}$ data; indeed it is a consequence of a more general interesting statement. Let's consider any $\alpha_1 < \alpha_2 < \alpha_3$, and $\beta_1  < \beta_2 < \beta_3$ with $\alpha_i \leq \beta_i$. Now, consider the set of all indices $\{1, 2, \cdots, n \}$ - $\{\alpha_1 + 1, \cdots, \alpha_3 - 1\}$ - $\{\beta_1 + 1, \cdots, \beta_3 - 1\}$, and choose any $a<b$ in this set. Then the claim is that 
\begin{equation}
{\rm det} \left| \begin{array}{cc} [a \alpha_1 \alpha_2 \alpha_3 ] & [b \beta_1 \beta_2 \beta_3 ] \cr [a \beta_1 \beta_2 \beta_3 ] & [b \alpha_1 \alpha_2 \alpha_3 ] \end{array} \right|>0
\end{equation}
These statement can be proven recursively, starting from ${\cal Z}$'s that correspond to $0$-dimensional cells of the external data positive Grassmannian where they are easily verified, and building up to a general configuration of ${\cal Z}$'s by shifting adjacent columns; it is easy to show that the shifts push all such determinants to be positive. 

Similarly for $k=3$, the analog of this statement is that for any $a<b<c$,\,$i<j<k$, we have 
\begin{equation}
{\rm det} \left| \begin{array}{ccc} [{\cal Z}_a {\cal Z}_b {\cal Z}_{i-1} {\cal Z}_i {\cal Z}_{i+1} ] & [ {\cal Z}_a {\cal Z}_b {\cal Z}_{j-1} {\cal Z}_j {\cal Z}_{j+1} ] & [{\cal Z}_a {\cal Z}_b {\cal Z}_{k-1} {\cal Z}_k {\cal Z}_{k+1} ] \cr [ {\cal Z}_a {\cal Z}_c {\cal Z}_{i-1} {\cal Z}_i {\cal Z}_{i+1} ] & [{\cal Z}_a {\cal Z}_c {\cal Z}_{j-1} {\cal Z}_j {\cal Z}_{j+1} ] & [{\cal Z}_a {\cal Z}_c {\cal Z}_{k-1} {\cal Z}_k {\cal Z}_{k+1} ] \cr 
[{\cal Z}_b {\cal Z}_c {\cal Z}_{i-1} {\cal Z}_i {\cal Z}_{i+1} ] & [{\cal Z}_b {\cal Z}_c  {\cal Z}_{j-1} {\cal Z}_j {\cal Z}_{j+1} ] & [{\cal Z}_b {\cal Z}_c {\cal Z}_{k-1} {\cal Z}_k {\cal Z}_{k+1} ] \end{array} \right|>0 
\end{equation}
This follows from a more general statement where $(i-1, i, i+1), (j-1,j,j+1),(k-1,k,k+1)$ are replaced by any $\alpha_1<\alpha_2<\alpha_3; \beta_1<\beta_2<\beta_3; \gamma_1<\gamma_2<\gamma_3$ with $\alpha_i \leq \beta_i \leq \gamma_i$, and $a,b,c$ are chosen from the set $\{1,\cdots, n\} - \{\alpha_1 + 1, \cdots, \alpha_3 - 1\} -  \{\beta_1 + 1, \cdots, \beta_3 - 1\}  -  \{\gamma_1 + 1, \cdots, \gamma_3 - 1\}$ with $a<b<c$.  The obvious generalization of these statements holds for higher $k$.  

The extension of the inequalities cutting out the tree amplituhedron to any $m$ is straightforward. For instance for $m=4$, we consider the 5-planes $({\cal Z}_{a-2} {\cal Z}_{a-1} {\cal Z}_a {\cal Z}_{a+1} {\cal Z}_{a+2})$, again multiplied by appropriate factors of $(-1)^{k-1}$ for indices that wrap past $n$. These 5-planes intersect the $k$-plane $Y$ in points $U_a$. Once again, we conjecture that $Y$ is in the amplituhedron if and only if the obvious boundaries $[Y i i+1 jj+1] > 0$, and the configuration of $k$-dimensional vectors $U_a$ is in $G_+(k,n)$. The extension to the all-loop amplituhedron then follows. We have $[ Y ii+1 jj+1 ]> 0, [ (Y A B)_\alpha i i+1 ] > 0, [Y (AB)_\alpha (AB)_\beta] > 0$. We also demand that the 3-planes $({\cal Z}_{a-1} {\cal Z}_a {\cal Z}_{a+1})$ intersect the $(k+2)$-planes $(YAB)_\alpha$ in points $V^{\alpha}_a$ which are belong to $G_+(k+2,n)$, and the 5-planes $({\cal Z}_{a-2} {\cal} Z_{a-1} {\cal Z}_a {\cal Z}_{a+1} {\cal Z}_{a+2})$ intersect $Y$ in points $U_a$ which belong to $G_+(k,n)$.

\section{Open Problems and Outlook}

We have presented an essentially combinatorial/topological characterization of the amplituhedron. It is remarkable that the rich, intricate geometry of the amplituhedron, and associated with it, all the non-trivial physics of planar ${\cal N}=4$ SYM scattering amplitudes, can ultimately determined by nothing more than specifying a simple pattern of $+$ and $-$ sign flips. 

A great deal remains to be understood both about the mathematics and physics associated with this new picture. Most pressingly, we would like to fully establish the equivalence of our new formulation of the amplituhedron with the usual one; all that remains to be shown is that satisfying correct winding or flip patterns implies that $Y$ can be written in the  ``$Y = C \cdot {\cal Z}$" form. At an even more basic level, we would like to have a proof of the equivalence between  ``sign-flip" and ``winding/crossing number" pictures. 

We have largely focused on describing points on the interior of the amplituhedron, but it is desirable to find an  characterization of all the boundaries of the amplituhedron along the same lines. 
On boundaries of the amplituhedron, many of the brackets involving $Y$ vanish and, for instance, the ``sign flip" criterion becomes ill-defined. We can of course ask if there are perturbations to $Y$ that change $``0"$'s into $+'s$ and $-$'s to get the right pattern of sign flips, but is there a more efficient combinatorial check of whether degenerate configurations of $Y$'s are in fact legal boundaries of the amplituhedron? 

For the simplest $m=1$ and $m=2$ amplituhedra, we saw that an exhaustive account of the sign flip/winding patterns directly led to triangulations of the spaces and the determination of their associated canonical forms. This picture does not trivially extend to higher $m$, but is there any topological interpretation of the known triangulations of $m=4$ amplituhedra, and if not, are there new triangulations that are more natural from the ``winding/flip" point of view? 

Do the sectors with different winding numbers have role to play in the physics? We have seen that the space defined purely by the obvious physical inequalities, even further generalized to simply mutual positivity between the 2-planes defining loops, seems to be related to ${\cal M} \times \overline{\cal M}$ and correlation functions; but what is the invariant property of the canonical form 
generalizing the notion of ``logarithmic singularities" which can determine correlators from the geometry? 

Finally, the $m$ dimensional image of the amplituhedron made possible by our new picture seems important from a number of points of view. In one obvious direction, we can finally treat the geometry of ``the integrand" and ``the amplitude" on exactly the same footing (see also \cite{Dixon:2016apl}). This should be especially useful in the context of the powerful new methods being developed, using the amplituhedron together with Landau equations, to constrain (and perhaps determine) the ``symbol" of multiloop MHV amplitudes in ${\cal N}=4$ SYM \cite{Goncharov:2010jf,Golden:2013xva,Dennen:2016mdk}. The winding/flip picture of the amplituhedron should reduce this program to perfectly well-defined geometry problems, not just for MHV amplitudes but for amplitudes with all $n,k,L$. 

It is also exciting to have a new picture  of the integrand of scattering amplitudes, which depending {\it solely} on the physical (momentum-twistor) data determining the momenta of the particles.  We have seen that $4 \times k$ forms on this kinematical space, which have logarithmic singularities on regions with correct winding numbers, determine the maximally supersymmetric amplitudes. It would be fascinating to extend this picture to the other examples of amplitudes which are known to be connected to positive geometry---for instance in ordinary momentum space (or ordinary twistor space) for ${\cal N}=4$ SYM, where ``winding" should plausibly make contact with twistor-strings \cite{Witten:2003nn}, and ABJM theory \cite{Aharony:2008ug,Huang:2013owa}.  

But more ambitiously, the notion of combining all helicity information together in one object as a {\it differential form}, rather than exploiting polarization vectors, or using the ``$\eta$"'s of supersymmetric theories, and fixing this form by singularities determined by topological properties, is a simple and powerful idea that begs for generalization. Since everything now depends only on the momenta of external particles,  our geometric, topological and combinatorial imaginations are no longer necessarily shackled to toy worlds with conformal invariance and supersymmetry, and we can hope to describe scattering amplitudes closer to the real world in this language. As some first steps in this direction, we are naturally led to ask: what happens when we have additional data, like lines at infinity that break conformal invariance; are there new notions of ``winding" associated with these structures? And it is peculiar to restrict our attention to $4 \times k$ forms only, are there natural forms of all degrees, which would certainly be associated with less (or non)supersymmetric theories?

\section*{Acknowledgements} We would like to thank Yuntao Bai, Song He, Thomas Lam, Steve Karp and Lauren Williams for useful discussions. The work of NAH is supported by the DOE under grant  DOE DE-SC0009988.  The work of HT is supported by NSERC and the Canada Research Chairs program.  We gratefully recognize the hospitality of the Fields Institute, where some of this work was carried out.  


\begin{thebibliography}{99}

\bibitem{ArkaniHamed:2012nw}
  N.~Arkani-Hamed, J.~L.~Bourjaily, F.~Cachazo, A.~B.~Goncharov, A.~Postnikov and J.~Trnka,
  {\it Cambridge University Press}, arXiv:1212.5605 [hep-th].

\bibitem{L1}
G.~Lusztig, Representation Theory {\bf 2} (1998)  70--78.

\bibitem{L2}
G.~Lusztig, Lie Theory and
  Geometry, In Honor of B. Kostant, vol.~123 of Prog. in Math.,
  pp.~531--569. Birkhauser, Boston, 1994.

\bibitem{P}
A.~Postnikov, arXiv:math/0609764.
  
\bibitem{Arkani-Hamed:2013jha}
  N.~Arkani-Hamed and J.~Trnka,
  JHEP {\bf 1410}, 30 (2014)
  [arXiv:1312.2007 [hep-th]].

\bibitem{Arkani-Hamed:2013kca}
  N.~Arkani-Hamed and J.~Trnka,
   JHEP {\bf 1412}, 182 (2014)
 [arXiv:1312.7878 [hep-th]].

  \bibitem{Franco:2014csa}
  S.~Franco, D.~Galloni, A.~Mariotti and J.~Trnka,
  JHEP {\bf 1503}, 128 (2015)
   [arXiv:1408.3410 [hep-th]].
 
  \bibitem{Bai:2014cna}
  Y.~Bai and S.~He,
  JHEP {\bf 1502}, 065 (2015)
  [arXiv:1408.2459 [hep-th]].
 
\bibitem{Lam:2014jda}
  T.~Lam,
  Commun.\ Math.\ Phys.\  {\bf 343}, no. 3, 1025 (2016)
  [arXiv:1408.5531 [math.AG]].

\bibitem{Arkani-Hamed:2014dca} 
  N.~Arkani-Hamed, A.~Hodges and J.~Trnka,
  JHEP {\bf 1508}, 030 (2015)
  [arXiv:1412.8478 [hep-th]].

\bibitem{Ferro:2015grk} 
  L.~Ferro, T.~Lukowski, A.~Orta and M.~Parisi,
  JHEP {\bf 1603}, 014 (2016)
  [arXiv:1512.04954 [hep-th]].
  
\bibitem{Galloni:2016iuj} 
  D.~Galloni,
  arXiv:1601.02639 [hep-th].

\bibitem{Ferro:2016zmx} 
  L.~Ferro, T.~Lukowski, A.~Orta and M.~Parisi,
  arXiv:1612.04378 [hep-th].

\bibitem{Ferro:2016ptt} 
  L.~Ferro, T.~Lukowski, A.~Orta and M.~Parisi,
  arXiv:1612.06276 [hep-th].

\bibitem{Karp:2016uax} 
  S.~N.~Karp and L.~K.~Williams,
  arXiv:1608.08288 [math.CO].

\bibitem{Bai:2015qoa} 
  Y.~Bai, S.~He and T.~Lam,
  JHEP {\bf 1601}, 112 (2016)
  doi:10.1007/JHEP01(2016)112
  [arXiv:1510.03553 [hep-th]].

\bibitem{Arkani-Hamed:2017tmz} 
  N.~Arkani-Hamed, Y.~Bai and T.~Lam,
  arXiv:1703.04541 [hep-th].

\bibitem{Hodges:2009hk}
  A.~Hodges,
  JHEP {\bf 1305}, 135 (2013)
  [arXiv:0905.1473 [hep-th]].

\bibitem{Goncharov:2010jf} 
  A.~B.~Goncharov, M.~Spradlin, C.~Vergu and A.~Volovich,
  Phys.\ Rev.\ Lett.\  {\bf 105}, 151605 (2010)
  [arXiv:1006.5703 [hep-th]].

\bibitem{Golden:2013xva} 
  J.~Golden, A.~B.~Goncharov, M.~Spradlin, C.~Vergu and A.~Volovich,
  JHEP {\bf 1401}, 091 (2014)
  [arXiv:1305.1617 [hep-th]].


 
 \bibitem{Eden:2017fow} 
  B.~Eden, P.~Heslop and L.~Mason,
  arXiv:1701.00453 [hep-th].
   
\bibitem{Dixon:2016apl} 
  L.~J.~Dixon, M.~von Hippel, A.~J.~McLeod and J.~Trnka,
  JHEP {\bf 1702}, 112 (2017)
  [arXiv:1611.08325 [hep-th]].

\bibitem{Dennen:2016mdk} 
  T.~Dennen, I.~Prlina, M.~Spradlin, S.~Stanojevic and A.~Volovich,
  arXiv:1612.02708 [hep-th].

\bibitem{Witten:2003nn} 
  E.~Witten, Commun.\ Math.\ Phys.\  {\bf 252}, 189 (2004), hep-th/0312171.

\bibitem{Aharony:2008ug} 
  O.~Aharony, O.~Bergman, D.~L.~Jafferis and J.~Maldacena,
  JHEP {\bf 0810}, 091 (2008)
  [arXiv:0806.1218 [hep-th]].

\bibitem{Huang:2013owa} 
  Y.~Huang, C.~Wen,  JHEP (2014) 2014: 104. arXiv:1309.3252 [hep-th].




\end{thebibliography}
\end{document}